\documentclass[prl,aps,amssymb,twocolumn,superscriptaddress, notitlepage, longbibliography]{revtex4-2}

\usepackage{bm,amssymb,amsmath,dsfont,mathrsfs,amstext,latexsym,physics,relsize}
\usepackage[dvipsnames]{xcolor}
\usepackage[colorlinks=true,citecolor=RoyalBlue,urlcolor=RoyalBlue,linkcolor=RoyalBlue]{hyperref}
\usepackage{orcidlink}

\newcommand{\rev}[1]{{#1}}

\begin{document}

\title{Shortcuts to Adiabaticity via Adaptive Quantum Zeno Measurements}

\author{Adolfo~del Campo\orcidlink{0000-0003-2219-2851}}
\email{adolfo.delcampo@uni.lu}
\affiliation{Department  of  Physics  and  Materials  Science,  University  of  Luxembourg,  L-1511  Luxembourg, Luxembourg}
\affiliation{Donostia International Physics Center,  E-20018 San Sebasti\'an, Spain}

\begin{abstract}
Stroboscopic measurements of a time-dependent projector generates a quantum Zeno dynamics whose effective Hamiltonian is the Kato-Avron Hamiltonian for parallel transport, which reduces to counterdiabatic driving in the case of a nonselective measurement of the instantaneous eigenprojectors of the system. The same generator arises under continuous monitoring of a time-dependent observable and under a time-varying complex absorbing potential. Departures from the ideal Zeno limit are governed, in all three settings, by a single positive leakage operator with a dynamical and a purely geometric contribution. Matching the corresponding generators yields a model-independent generalized Schulman relation for the quantum Zeno dynamics in the three scenarios, and a purely geometric bound on the measurement resources needed to steer a subspace along a prescribed path.
\end{abstract}

\maketitle

Stirring the dynamics of a quantum system towards a target state or subspace is a ubiquitous task in quantum science and technology. Adiabatic protocols provide a natural approach to this end. However, they require slow driving, which can lead to the accumulation of uncontrollable sources of noise and errors.  Shortcuts to adiabaticity offer an alternative, achieving the same target state under fast driving \cite{Chen2010,Torrontegui2013,GueryOdelin2019,Hatomura2024}. Counterdiabatic driving (CD) \cite{Demirplak2003,Demirplak2005,Demirplak2008}, also known as transitionless quantum driving \cite{Berry2009}, relies on auxiliary controls to fast-forward the adiabatic evolution, and has been generalized to accelerate an arbitrary reference trajectory \cite{Alipour2020}.
At the single-particle level, it has been experimentally demonstrated in nitrogen-vacancy centers \cite{Zhang2013}, trapped ions \cite{An2016}, atomic ensembles \cite{Du2016}, and superconducting qubits \cite{Wang2018}, among others.
Its realization in many-body systems is challenging, as the CD term involves spatially nonlocal and multiple-body interactions \cite{delCampo2012}. This has motivated a variety of approximation schemes, including variational methods \cite{Saberi2014,Sels2017} and expansions with controlled nonlocality \cite{Claeys2019,Takahashi2023,Grabarits2025}, giving rise to a family of heuristic quantum algorithms for quantum optimization \cite{Chandarana22,Hegade22} demonstrated via digital quantum simulation \cite{Zhou2020,Hegade2021,Visuri2025,Hatomura2026}.
We note that CD is closely related to parallel transport, which describes a driven system in which the probability amplitude in a given basis remains constant in time up to a phase factor. The generator of parallel transport has been known since early proofs of the adiabatic theorem and its generalizations \cite{Kato1950,Avron1999,Teufel2003}.

CD techniques rely on coherent quantum control, although they can be generalized to open quantum systems with nonunitary dynamics \cite{Vacanti2014,Alipour2020}.
Other prominent approaches to control quantum evolution utilize quantum measurements.
An instance is the celebrated quantum Zeno effect, in which frequent projective measurements freeze the time evolution and the coupling to the surrounding environment \cite{Misra1977,Itano1990,Facchi2004}: the evolution of a system with Hamiltonian $H$, subject to frequent measurements of a projector $P$, is described by the effective Hamiltonian $H_{\rm Z}=PHP$ in the Zeno limit \cite{Facchi2002,Facchi2008}.
The extension to adaptive quantum measurements has broad applications \cite{AharonovVardi80,Pechen06,Jacobs10,Tanaka2012,Burgarth2013,Lewalle2024}. 
Adaptive Zeno dynamics has been proposed as an alternative to adiabatic transport in the study of geometric phases \cite{Burgarth2013}.
The quantum Zeno effect has also been proposed \cite{Pechen06,Tanaka2012} and used experimentally to guide the evolution of a quantum system \cite{Blok2014,Hacohen-Gourgy2018} via adaptive and continuous measurements of a time-varying observable, an approach whose analogy with CD has been pointed out \cite{Lewalle2024}. 
Adaptive Zeno measurements may also be used to engineer quantum heat engines \cite{Barontini25} and for ground state preparation \cite{Kuo2021,Watts2026}.

In this Letter, we derive the effective Hamiltonian under a sequence of time-varying projective measurements. This result generalizes the celebrated expression for quantum Zeno dynamics by including an additional geometric term, the Kato-Avron Hamiltonian that generates parallel transport. For projective measurements of the instantaneous energy eigenbasis of the driven quantum system, the quantum Zeno Hamiltonian becomes identical to that under CD, while the global time evolution is nonunitary and suppresses quantum coherence among the measurement eigenspaces. These results are generalized to continuous quantum measurements with a time-varying observable, and to non-Hermitian quantum evolution with a time-dependent complex absorbing potential. Away form the Zeno limit, we compute the leading non-Hermitian correction for all three protocols and show that a single positive leakage operator controls them, with a dynamical and a purely geometric contribution. This fixes an exact operator-level correspondence between pulsed and continuous adaptive monitoring and yields a geometric lower bound on the leakage, and on the measurement resources, of any Zeno-dragging protocol. Thus, adaptive quantum Zeno measurements provide a versatile approach to engineer shortcuts to adiabaticity.

{\it Stroboscopic Zeno dynamics with an adaptive projector.---}
Consider a quantum system with Hamiltonian $H(t)$ acting on a Hilbert space $\mathcal{H}$ \cite{AharonovVardi80,Burgarth2013}. We perform a sequence of projective measurements of a time-dependent projector $P(t)$ at times $t_k = k\,\delta t$ for $k=0,\dots,N$, interspersed with unitary evolution generated by $H(t)$. Let us assume that $P(t)$ is differentiable in time, $P(t)^2=P(t)$, $P(t)^\dagger=P(t)$, and of constant rank.
The stroboscopic evolution operator is
\begin{equation}
U_N(T)=\prod_{k=0}^{N-1} P(t_{k+1})e^{-iH(t_k)\delta t}P(t_k),
\end{equation}
with $T=N\delta t$ and time ordering implicit.
Expanding to first order in $\delta t$,
$e^{-iH(t_k)\delta t} = \mathbb{I} - iH(t_k)\delta t$ and
$P(t_{k+1}) = P(t_k) + \dot P(t_k)\delta t$, each factor becomes
\begin{eqnarray}
& & P(t_{k+1})e^{-iH(t_k)\delta t}P(t_k) = \bigl(P + \dot P\delta t\bigr)
\bigl(\mathbb{I}-iH\delta t\bigr)P  \nonumber\\
&&= P - iPHP\delta t + \dot P P\delta t ,
\end{eqnarray}
where all operators are evaluated at $t_k$.
Further note that $P\dot P P = 0$ \cite{Kato1950}, which follows from differentiating $P^2=P$, and thus
$\dot P P = [\dot P,P]P$. Ignoring $\mathcal{O}(\delta t^2)$ corrections, it follows that
\begin{equation}
P(t_{k+1})e^{-iH\delta t}P(t_k)=\left[\mathbb{I} - i\delta t\left(PHP + i[\dot P,P]\right)\right]P.
\end{equation}
The quantum Zeno dynamics arises in the limit $N\to\infty$, $\delta t\to 0$ with the total time of evolution $T$ fixed. The product formula converges to a time-ordered exponential acting within the instantaneous subspace,
$U_{\rm Z}(T)=\mathcal{T}\exp\left(-i\int_0^T H_{\mathrm{Z}}(t)dt\right)P(0)$,
where the effective Hamiltonian is
\begin{equation}
H_{\rm Z}(t)=P(t)H(t)P(t)+i[\dot P(t),P(t)].
\label{KatoAvronH}
\end{equation}
The first term is the projected physical Hamiltonian, while the second term is a geometric contribution associated with parallel transport in the moving subspace associated with $P(t)$. Indeed, this is precisely the Kato-Avron Hamiltonian introduced in proofs of the adiabatic theorem \cite{Kato1950,Avron1999,Teufel2003}.

This result can be readily generalized to a complete orthogonal set of projectors $\{P_n(t)\}$ satisfying
$P_n(t)P_m(t)=\delta_{nm}P_n(t)$ and
$\sum_n P_n(t)=\mathbb{I}$. Measuring a full set of outcomes $\{P_n(t)\}$ at each step yields a projection-valued measure (PVM), for which a similar analysis yields
\begin{equation}
H_{\rm Z}(t)=
\sum_n P_n(t)H(t)P_n(t)
+
\frac{i}{2}\sum_n[\dot P_n(t),P_n(t)].
\label{Heff_general}
\end{equation}
\rev{The factor $\tfrac{1}{2}$ arises because $S\equiv\sum_n\dot P_nP_n$ is anti-Hermitian, so that $\sum_n[\dot P_n,P_n]=2S$ \cite{SM}; for the two-outcome PVM $\{P,Q\}$ Eq.~(\ref{Heff_general}) reduces to Eq.~(\ref{KatoAvronH}).}
The effective Hamiltonian is block-diagonal in the instantaneous decomposition of the Hilbert space, with $P_n H P_n$ generating the dynamics within each sector, while the connection term generates parallel transport in each subspace. Transitions between sectors are thus suppressed in the Zeno limit.

\rev{The leading correction for a finite measurement interval $\delta t$ induces a non-Hermitian term in the effective Hamiltonian,
\begin{equation}
H_{\mathrm{eff}}^{(\delta t)}(t)=H_{\rm Z}(t)-\frac{i\delta t}{2}\,\Gamma(t)+\mathcal{O}(\delta t^2),
\label{eq:Heffdt}
\end{equation}
where the leakage operator reads
\begin{equation}
\Gamma(t)=P\big(H-A\big)Q\big(H-A\big)P,
\qquad
A(t)\equiv i[\dot P,P],
\label{eq:Gammadef}
\end{equation}
written for the two-outcome PVM $\{P,Q\}$. For a general PVM, one has $A=\tfrac{i}{2}\sum_n[\dot P_n,P_n]$ and $\Gamma=\sum_n P_n(H-A)(\mathbb I-P_n)(H-A)P_n$. Explicitly,
\begin{equation}
\Gamma=PHQHP+P\dot P\dot PP-i\big(PHQ\dot PP-P\dot PQHP\big).
\label{eq:Gammaexp}
\end{equation}
The first contribution is the familiar static-projector leakage induced by the coupling between the Zeno subspace and its complement. It is the only surviving term for $\dot P=0$, in which case Eq.~(\ref{eq:Heffdt}) reduces to the static result $H_{\rm eff}=PHP-\tfrac{i\delta t}{2}PHQHP$. The second contribution is purely geometric and quantifies the leakage caused by the motion of the projector itself. It is the only surviving term when $P(t)$ is a spectral projector of $H(t)$, for which $QHP=0$. The cross term requires both $QHP\neq0$ and $\dot P\neq0$. Note that $\Gamma$ is a positive operator, $\Gamma=X^\dagger X$ with $X=Q(H-A)P$, so its anti-Hermitian contribution can only reduce the norm.
The derivation is given in the End Matter and is most transparent in the co-moving frame that freezes the projector, where the problem maps exactly onto a static-projector Zeno problem with the transformed Hamiltonian $\tilde H=W^\dagger(H-A)W$.}

{\it Relation to Counterdiabatic Driving.---}
When the monitored projectors are chosen to be the instantaneous eigenprojectors of the system Hamiltonian, the effective Zeno Hamiltonian $H_{\rm Z}(t)$ reduces to the CD Hamiltonian, widely used for shortcuts to adiabaticity \cite{Demirplak2003,Demirplak2005,Demirplak2008,Berry2009}.
Let the system Hamiltonian admit the spectral decomposition
$H(t)=\sum_n E_n(t)P_n(t)$,
with orthogonal and differentiable projectors $P_n(t)$. The first term in Eq.~(\ref{Heff_general}) then collapses onto the spectral decomposition, $\sum_n P_n H P_n=H(t)$, while the additional contribution $A$ is purely geometric and depends only on the motion of the spectral projectors. For a nondegenerate spectrum, with unit-rank projectors $P_n(t)=|n(t)\rangle\langle n(t)|$, differentiation gives $\dot P_n=|\dot n\rangle\langle n|+|n\rangle\langle \dot n|$, and summing over $n$ using completeness \cite{SM} one finds
\begin{equation}
H_{\rm Z}=H(t)+
i\sum_n\left(|\dot n\rangle\langle n|-\langle n|\dot n\rangle |n\rangle\langle n|\right).
\label{At_Berryform}
\end{equation}
Equation (\ref{At_Berryform}) coincides with the CD Hamiltonian derived in the theory of transitionless quantum driving \cite{Demirplak2003,Demirplak2005,Demirplak2008,Berry2009}, which generates exact adiabatic transport along the instantaneous eigenstates of $H(t)$ without requiring slow driving.

Note that both the adiabatic evolution and the CD dynamics are described by (entropy-preserving) unitaries that maintain quantum coherence in the energy eigenbasis. By contrast, the Zeno implementation based on frequent nonselective measurements of $\{P_n(t)\}$ generates a quantum channel of the form
$\rho(t)=\sum_n U_n(t)\,P_n(0)\,\rho_0\,P_n(0)\,U_n^\dagger(t)$,
where $U_n(t)$ are unitary operators acting within the corresponding subspaces. The measurement-induced pinching map
$\Lambda(\rho_0)=\sum_n P_n(0)\,\rho_0\,P_n(0)$ eliminates coherences between different sectors. Such pinching operations are purity-nonincreasing and entropy-nondecreasing, with equality holding only when $\rho_0$ is already block-diagonal in the decomposition defined by $\{P_n(0)\}$. Thus, although the quantum Zeno dynamics reproduces the same geometric generator responsible for CD in each instantaneous eigenspace, it realizes this transport through an incoherent, entropy-producing process: an initial pure state $|\psi(0)\rangle=\sum_nc_n(0)|n(0)\rangle$ evolves under adaptive Zeno measurements into the mixed state $\rho(t)=\sum_n|c_n(0)|^2P_n(t)$, rather than into the coherent superposition generated by CD \cite{SM}.
\rev{This dephasing across sectors is a distinguishing feature from conventional CD and common to other measurement-induced quantum control approaches \cite{Roa06,Pechen06,Jacobs10,Blok2014,Hacohen-Gourgy2018}. Whenever the initial state is block diagonal in $\{P_n(0)\}$, exact CD and the adaptive Zeno protocol implement the same channel; they differ only on initial coherences across sectors, which the Zeno protocol elliminates. The Zeno route is therefore preferable when such coherences are irrelevant or unwanted, as in driven thermodynamic processes \cite{Funo2017,Barontini25}, where a coherence-free trajectory in the instantaneous energy basis removes the coherent contribution to the work statistics, or in sector-resolved state preparation, as further discussed in \cite{SM}.
}

{\it Shortcuts via continuous monitoring.---}
We next consider continuous quantum measurements \cite{WisemanMilburn2010,Jacobs2014} and show that the same Zeno Hamiltonian emerges from the stochastic master equation (SME) in the strong-measurement limit. Consider the continuous monitoring of a single Hermitian observable
$X(t)=\sum_n x_n\,P_n(t)$,
with nondegenerate eigenvalues $\{x_n\}$, which is operationally equivalent to monitoring the subspaces $\mathrm{Ran}\,P_n(t)$. The diffusive (homodyne-type) SME for the conditioned state $\rho_c(t)$ reads
\begin{eqnarray}
d\rho_c
& =&
-i[H(t),\rho_c]\,dt
+\kappa\,\mathcal D[X(t)]\,\rho_c\,dt\nonumber\\
& & +\sqrt{\kappa}\,\mathcal H[X(t)]\,\rho_c\,dW(t),
\label{eq:SME-X}
\end{eqnarray}
where $dW$ is a Wiener increment ($\mathbb E[dW]=0$, $dW^2=dt$), $\kappa$ is the measurement strength, $\mathcal D[X]\rho=X\rho X-\tfrac12\{X^2,\rho\}$ and $\mathcal H[X]\rho=\{X,\rho\}-2\langle X\rangle\rho$. Averaging Eq.~\eqref{eq:SME-X} over the measurement outcomes yields the unconditional master equation $\dot\rho=-i[H,\rho]+\kappa\,\mathcal D[X]\,\rho$, which describes measurement-induced dephasing in the instantaneous eigenspaces of $X(t)$, i.e., the PVM $\{P_n(t)\}$.

To connect with the stroboscopic Zeno Hamiltonian, it is useful to introduce the unitary intertwiner $W(t)$ that transports the PVM,
$\dot W=K W$ with $K\equiv -iA=\tfrac{1}{2}\sum_n[\dot P_n,P_n]$ and $W(0)=\mathbb I$, for which $P_n(t)=W(t)P_n(0)W^\dagger(t)$.
Because $W(t)$ is deterministic, the It\^o transformation of the co-moving conditioned state $\tilde\rho_c=W^\dagger\rho_cW$ gives an SME of the same form, but with time-independent measurement operators $X_0\equiv X(0)$ and the transformed Hamiltonian $\tilde H=W^\dagger H W-i\,W^\dagger\dot W$ \cite{SM}.
Thus, the same operator that generated the geometric term in the stroboscopic derivation appears here as the inertial (connection) term $-iW^\dagger\dot W$ in the co-moving frame; see also \cite{Lewalle2024}.
The Zeno regime corresponds to $\kappa$ being the largest rate in the problem, for which the dissipator $\kappa\mathcal D[X_0]$ rapidly damps the off-diagonal blocks $P_n(0)\tilde\rho_c P_m(0)$ with $n\neq m$. As $\kappa\to\infty$, the dynamics becomes confined to the block-diagonal manifold and the remaining evolution within each block is generated, to leading order in $1/\kappa$, by $P_n(0)\tilde H(t)P_n(0)$.
Transforming back to the laboratory frame yields an effective unitary evolution within the moving subspaces with the Hamiltonian (\ref{Heff_general}), in agreement with the stroboscopic Zeno result. For spectral projectors of $H(t)$ one may choose $W(t)=\sum_n|n(t)\rangle\langle n(0)|$ and the effective Hamiltonian equals the CD Hamiltonian, while the two-outcome choice $X=P(t)$ yields Eq.~(\ref{KatoAvronH}).

{\it Shortcuts via time-dependent complex absorbing potentials.---}
Complex absorbing potentials have long been used to model detectors, e.g., in the time of arrival to a given quantum subspace \cite{Allcock1969a,Allcock1969b,Allcock1969c,Muga2004}, one of the earliest manifestations of the quantum Zeno effect, predating the celebrated work of Misra and Sudarshan \cite{Misra1977}. Continuous and pulsed observations can thus both induce the quantum Zeno effect \cite{Schulman1998,Streed2006,Facchi01,Muga2008}. As we next show, time-dependent complex absorbing potentials give rise, in the Zeno limit, to an effective Kato-Avron Hamiltonian, and to the CD Hamiltonian as a special case. Consider the non-Hermitian Hamiltonian
\begin{equation}
H_{\rm nh}(t)=H(t)-i\kappa\,Q(t),
\label{eq:Hnh}
\end{equation}
where $H(t)=H^\dagger(t)$ is the physical Hamiltonian, $Q(t)$ is a projector, and $\kappa>0$ sets the strength of the absorbing potential. The complementary projector $P(t)=\mathbb I-Q(t)$ defines the Zeno subspace: for large $\kappa$, components of the wavefunction in $\mathrm{Ran}\,Q(t)$ decay rapidly, so that the long-time evolution should remain confined to $\mathrm{Ran}\,P(t)$.
Working again in the co-moving frame $|\tilde\psi\rangle=W^\dagger|\psi\rangle$ defined by the intertwiner $\dot W=KW$, $K=[\dot P,P]$, the projectors become time independent, $P(t)=WP_0W^\dagger$, and the Schr\"odinger equation reads $i\partial_t|\tilde\psi\rangle=(\tilde H-i\kappa Q_0)|\tilde\psi\rangle$ with $\tilde H=W^\dagger H W-iW^\dagger\dot W$. Decomposing $|\tilde\psi\rangle=|\tilde\psi_P\rangle+|\tilde\psi_Q\rangle$ and adiabatically eliminating the overdamped absorbing sector, $\dot{\tilde\psi}_Q\simeq0$, gives $|\tilde\psi_Q\rangle\simeq(i\kappa)^{-1}Q_0\tilde HP_0|\tilde\psi_P\rangle$ and hence \cite{SM}
\begin{equation}
\tilde H_{\rm eff}(t)
=
P_0\tilde HP_0
-\frac{i}{\kappa}\,
P_0\tilde HQ_0\tilde HP_0
+\mathcal{O}(\kappa^{-2}).\nonumber
\end{equation}
The first term is Hermitian and generates coherent dynamics within the Zeno subspace, while the second is anti-Hermitian and describes residual loss due to finite absorption strength.
Transforming back to the laboratory frame, the leading term reproduces the Kato-Avron Hamiltonian (\ref{KatoAvronH}), while the first correction in $1/\kappa$ gives
\begin{equation}
H_{\rm eff}^{(\kappa)}(t)
=
P H P
+
i[\dot P,P]-\rev{\frac{i}{\kappa}}\,\Gamma(t)+\mathcal{O}(\kappa^{-2}),
\label{KatoAvronviaNH}
\end{equation}
\rev{with the same positive leakage operator $\Gamma(t)$ of Eq.~(\ref{eq:Gammadef}), which here quantifies the leakage induced by a finite absorption strength.} 
%Note the prefactor: $-i/\kappa$ for the absorber, against $-i\delta t/2$ for pulsed measurements. This asymmetry is physical and fixes the correspondence between the two protocols, as discussed below.}
As before, if $P(t)$ is chosen to be a spectral projector of $H(t)$ (or a sum thereof), the projected term satisfies $P(t)H(t)P(t)=H(t)$ within the corresponding eigenspace, and the geometric term coincides with the CD Hamiltonian that enforces adiabatic transport \cite{SM}.
Thus, a strong time-dependent absorbing potential provides a dynamical realization of transitionless evolution that is formally equivalent, at leading order, to both stroboscopic Zeno measurements and shortcuts via continuous monitoring.

{\it Generalized Schulman relation.---}
In the static case, Schulman established a relation between the Zeno dynamics induced by stroboscopic measurements and complex absorbing potentials \cite{Schulman1998,Facchi01,Muga2008}, which was experimentally verified \cite{Streed2006}. Such a connection was established by analyzing the response time of a two-level system, leading to a relation between $\delta t$ and $\kappa$. This relation can be generalized by comparing the generators of evolution, rather than the response time scale, in a model-independent way for time-dependent Zeno measurements. \rev{Specifically, the two effective generators share one and the same leakage operator \cite{SM},
$H_{\rm eff}^{(\delta t)}=H_{\rm Z}-\tfrac{i\delta t}{2}\Gamma$ and $H_{\rm eff}^{(\kappa)}=H_{\rm Z}-\tfrac{i}{\kappa}\Gamma$,
with $\Gamma$ given by Eq.~(\ref{eq:Gammadef}). The two generators agree, to leading order in the respective small parameter, if and only if
\begin{equation}
\delta t=\frac{2}{\kappa},
\label{eq:Schulman}
\end{equation}
that is, the measurement interval equals twice the inverse absorption strength in the convention $H_{\rm nh}=H-i\kappa Q$. The factor of two has a transparent origin. As $\kappa$ is the decay rate of the amplitude in $\mathrm{Ran}\,Q$, the decay rate of the coherence between the two sectors is $\gamma_{\rm dec}=\kappa$, while probabilities decay at $2\kappa$. Written in terms of $\gamma_{\rm dec}$ the correspondence becomes convention independent,
\begin{equation}
\delta t=\frac{2}{\gamma_{\rm dec}},
\label{eq:SchulmanUniversal}
\end{equation}
and applies as well to continuous monitoring, where homodyne detection of $X=\sum_nx_nP_n(t)$ at rate $\kappa$ damps the coherence between sectors $n$ and $m$ at $\gamma^{nm}_{\rm dec}=\tfrac{\kappa}{2}(x_n-x_m)^2$ \cite{SM}. Equation~(\ref{eq:SchulmanUniversal}) generalizes the Schulman relation to time-dependent Zeno measurements at the level of the generator of the evolution, in a model independent way, and encompassing the three Zeno scenarios we have analyzed.}

{\it Discussion.---}
In summary, quantum control via shortcuts to adiabaticity can be induced via adaptive quantum Zeno measurements
in three complementary approaches: i) pulsed measurements of a time-dependent projector, ii) continuous quantum measurements of a time-dependent observable, and iii) non-Hermitian evolution with a time-dependent complex absorbing potential. In all cases, the effective Zeno Hamiltonian yields the Kato-Avron Hamiltonian for parallel transport as a time-dependent geometric correction that steers the evolution within the varying Zeno subspace. When the Zeno measurements are associated with the spectral projectors of the instantaneous system Hamiltonian, the effective dynamics becomes identical to the CD evolution within each sector, while suppressing quantum coherences between different sectors \rev{\cite{Burgarth2013}}.
\rev{Beyond that limit, the three routes are controlled by a single positive leakage operator, which fixes the generalized Schulman correspondence (\ref{eq:SchulmanUniversal}) and yields a concrete resource count. For spectral projectors the leakage is purely geometric, and the number of measurements needed to keep the total leakage below $\epsilon$ while dragging a state along a path of Fubini-Study length $\mathcal L$ obeys $N\gtrsim\mathcal L^2/\epsilon$, independently of the protocol duration $T$ (see End Matter). This is the measurement counterpart of the statement that exact CD requires a control field whose norm is set by the same geometric speed \cite{Demirplak2008,delCampo2012,Funo2017,Campbell17}, and it makes explicit when the Zeno route is competitive, i.e.,  whenever the available monitoring rate exceeds $\sim\mathcal L^2/(\epsilon T)$ and the inter-sector coherences destroyed by the measurement are not part of the target map. In dissipative platforms with natural decay channels (atoms with loss, polariton systems, monitored qubits) the absorbing-potential route is particularly advantageous as it requires neither collecting measurements nor synthesis of auxiliary coherent control fields, only the tuning of existing rates. Conversely, when the monitored projectors are many-body spectral projectors, engineering the measurement may be as demanding as engineering the CD term itself, and establishing a resource advantage in that regime remains an open problem. Finally, the Kato-Avron generator applies to any moving subspace, including multi-rank ones, so that the finite-strength analysis presented here also quantifies the error budget of the Zeno holonomies of Ref.~\cite{Burgarth2013} within degenerate manifolds, of direct relevance to holonomic quantum computation \cite{Zanardi1999,SM}.}
Our results provide a unified description of shortcuts to adiabaticity enabled by quantum Zeno measurements, which can be used in combination with or as an alternative to coherent quantum control, and should be of broad interest in quantum science and technology.

{\it Acknowledgements.---}
It is a pleasure to acknowledge discussions with Kasturi Ranjan Swain and an anonymous referee. 
The author is grateful to Andrew Jordan, Kater Murch, and David Weld as organizers of ``Quantum Interactive Matter: Harnessing Measurement and Feedback for Novel Quantum States'' at the Aspen Center for PHYics, supported by the National Science Foundation grant PHY-2210452. We further thank Wolfgang Ketterle for insightful discussions.
This work is supported by the  Luxembourg National Research Fund under Grant No. C24/MS/18940482/STAOpen. 

\bibliography{STAQZ}

\newpage

\appendix

\section{Finite-time-step corrections to Zeno dragging}\label{AppFinite}
In the ideal (mathematical) Zeno limit, $\delta t\to 0$, stroboscopic monitoring of a time-dependent projector $P(t)$ yields the effective Kato-Avron Hamiltonian within the Zeno subspace. In experiments, however, the measurement interval $\delta t$ is finite. Here we compute the leading correction in $\delta t$, which is most transparent in the co-moving frame in which the projector is frozen. Let $W(t)$ be the Kato intertwiner,
\begin{equation}
\dot W(t)=K(t)W(t),\quad K(t)=[\dot P(t),P(t)],\quad W(0)=\mathbb I,
\end{equation}
so that $P(t)=W(t)P_0W^\dagger(t)$ and $Q(t)=W(t)Q_0W^\dagger(t)$, with $P_0\equiv P(0)$ and $Q_0\equiv Q(0)$. The generator $K$ is anti-Hermitian, hence $W$ is unitary, and $A\equiv i[\dot P,P]=iK$ is Hermitian.

The exact one-step operator conditioned on remaining in the monitored subspace is
$\Omega(t)=P(t+\delta t)\,U(t+\delta t,t)\,P(t)$,
where $U$ is the propagator generated by $H$. Because $\Omega$ maps $\mathrm{Ran}\,P(t)$ into the displaced subspace $\mathrm{Ran}\,P(t+\delta t)$, it cannot be written as an operator sandwiched between two copies of $P(t)$: doing so would discard the first-order contribution associated with the motion of the projector. Conjugation with the intertwiner removes this obstruction. Defining
$\tilde\Omega(t)\equiv W^\dagger(t+\delta t)\,\Omega(t)\,W(t)$
and using $W^\dagger(t)P(t)W(t)=P_0$ at all times, one obtains the exact identity
\begin{equation}
\tilde\Omega(t)=P_0\,\tilde U(t+\delta t,t)\,P_0,
\label{eq:tildeOmega}
\end{equation}
where $\tilde U(t',t)=W^\dagger(t')U(t',t)W(t)$ is generated by the transformed Hamiltonian
\begin{equation}
\tilde H(t)=W^\dagger H W-iW^\dagger\dot W=W^\dagger(t)\big(H(t)-A(t)\big)W(t).
\label{eq:Htilde-strobo}
\end{equation}
Equation~\eqref{eq:tildeOmega} is exact and states that, in the co-moving frame, stroboscopic monitoring of a moving projector under $H(t)$ is identical to stroboscopic monitoring of the fixed projector $P_0$ under $\tilde H(t)$. The intertwiners act in the product of steps,
$U_N(T)=W(T)\prod_{k}P_0\,\tilde U(t_{k+1},t_k)\,P_0$,
so every finite-$\delta t$ contribution may be derived in the frozen frame and transported back at the end. This construction was already used in Ref.~\cite{Burgarth2013} to describe non-Abelian phases in the Zeno limit, while here it allow us to identify the effective Hamiltonian up to first order in $\delta t$.

Write $\tilde U(t+\delta t,t)=\exp(-i\delta t\,\bar{\tilde H})$ with $\bar{\tilde H}=\tilde H(t+\delta t/2)+\mathcal O(\delta t^2)$ the Magnus average over the step, and split the intermediate resolution of the identity as $\mathbb I=P_0+Q_0$, so that $P_0\bar{\tilde H}^2P_0=(P_0\bar{\tilde H}P_0)^2+P_0\bar{\tilde H}Q_0\bar{\tilde H}P_0$. Expanding \eqref{eq:tildeOmega} to second order then gives
\begin{eqnarray}
\tilde\Omega(t)&=&P_0\Big[\mathbb I-i\delta t\,\tilde H_{\rm Z}-\frac{\delta t^2}{2}\tilde H_{\rm Z}^2\Big]P_0
-\frac{\delta t^2}{2}\tilde\Gamma+\mathcal O(\delta t^3)\nonumber\\
&=&P_0\exp\Big[-i\delta t\Big(\tilde H_{\rm Z}-\frac{i\delta t}{2}\tilde\Gamma\Big)\Big]P_0+\mathcal O(\delta t^3),
\label{eq:Kreorg}
\end{eqnarray}
with $\tilde H_{\rm Z}\equiv P_0\bar{\tilde H}P_0$ and $\tilde\Gamma\equiv P_0\bar{\tilde H}Q_0\bar{\tilde H}P_0$.
Both sides of \eqref{eq:Kreorg} are operators from $\mathrm{Ran}\,P_0$ to itself, and no first-order contribution is lost, the motion of the projector having been absorbed into $\tilde H$.

Going back to the laboratory frame and using $PAP=0$, $QAP=iQ\dot PP$ and $PAQ=-iP\dot PQ$, one has $W\tilde H_{\rm Z}W^\dagger=PHP$. Restoring the inertial term $+A$ generated by $W$ yields Eq.~(\ref{eq:Heffdt}),
\begin{equation}
H^{(\delta t)}_{\rm eff}(t)=\underbrace{PHP+i[\dot P,P]}_{H_{\rm Z}(t)}-\frac{i\delta t}{2}\,\Gamma(t)+\mathcal O(\delta t^2),
\label{eq:Heff-dt-EM}
\end{equation}
with the leakage operator of Eqs.~(\ref{eq:Gammadef})--(\ref{eq:Gammaexp}), where we used $P\dot PQ\dot PP=P\dot P\dot PP$, a consequence of the Kato identity $P\dot PP=0$ \cite{Kato1950}.
The $\mathcal O(\delta t)$ Hermitian difference between $\tilde H(t)$ and $\tilde H(t+\delta t/2)$ results form  discretization. It is removed by sampling $H_{\rm Z}$ at the midpoint of each interval and does not affect the anti-Hermitian part. By contrast, $-\tfrac{i\delta t}{2}\Gamma$ is convention independent. The same result follows from a direct expansion in the laboratory frame, which yields $\Omega^\dagger\Omega=P-\delta t^2\,\Gamma+\mathcal O(\delta t^3)$ \cite{SM}. Hence, for a normalized state at time $t$ inside the Zeno subspace, the leakage probability per step is
\begin{equation}
p_{\rm leak}(t)=
\delta t^2\,
\big\|Q\big(H-i[\dot P,P]\big)P|\psi(t)\rangle\big\|^2
=\delta t^2\langle\Gamma\rangle,
\label{eq:pleak}
\end{equation}
up to $\mathcal{O}(\delta t^3)$ corrections, which is generically $\mathcal O(\delta t^2)$ for a state initially confined to $\mathrm{Ran}\,P(t)$. The two descriptions are thus mutually consistent. Over $N=T/\delta t$ steps, Eq.~\eqref{eq:pleak} gives a norm-squared decay rate $\delta t\,\langle\Gamma\rangle$, precisely the rate produced by the anti-Hermitian part of the generator \eqref{eq:Heff-dt-EM}.
For a time-independent projector ($\dot P=0$), Eq.~\eqref{eq:pleak} reduces to the standard static-Zeno leakage $p_{\rm leak}\simeq \delta t^2\langle H Q H\rangle$ \cite{Facchi2008}. For spectral projectors $P(t)$ of $H(t)$, $QHP=0$ so that $\Gamma=P\dot P\dot PP$ is purely geometric.

\section{A geometric bound on the measurement rate}\label{AppBound}
Information geometry sets fundamental bounds on the performance of shortcuts to adiabaticity \cite{Gong2022Bounds}, and thus is expected to affect their engineering by Zeno measurements.
The purely geometric character of the leakage for spectral projectors, the case relevant to shortcuts to adiabaticity, can serve to quantify the resources for the protocol. For a nondegenerate target, $P(t)=|n(t)\rangle\langle n(t)|$, one has $Q\dot PP|n\rangle=|\dot n\rangle-\langle n|\dot n\rangle|n\rangle$ and thus
\begin{equation}
\langle\Gamma\rangle=\langle\dot n|\dot n\rangle-|\langle n|\dot n\rangle|^2\equiv g(t),
\end{equation}
the Fubini-Study metric (the real part of the quantum geometric tensor) evaluated along the path \cite{Provost1980Riemannian}. Summing \eqref{eq:pleak} over $N=T/\delta t$ steps, the total leakage is
\begin{eqnarray}
p^{\rm tot}_{\rm leak}\simeq\delta t\int_0^T g(t)\,dt.
\end{eqnarray}
By the Cauchy-Schwarz inequality, $\int_0^Tg\,dt\ge\mathcal L^2/T$, where
$\mathcal L=\int_0^T\sqrt{g(t)}\,dt$ 
is the Fubini-Study length of the path traced by the monitored subspace. Hence,
\begin{equation}
p^{\rm tot}_{\rm leak}\gtrsim\frac{\delta t\,\mathcal L^{2}}{T}=\frac{\mathcal L^{2}}{N},
\label{eq:geobound}
\end{equation}
with equality for constant-speed dragging. Equivalently, the number of measurements required to steer the subspace along a prescribed path with total leakage below $\epsilon$ satisfies
\begin{equation}
N\gtrsim\frac{\mathcal L^{2}}{\epsilon},
\end{equation}
regardless of the protocol duration $T$. The cost is set by the geometry of the path in state space, not by how fast it is traversed. Using the correspondence \eqref{eq:SchulmanUniversal}, the same bound reads 
\begin{eqnarray}
\gamma_{\rm dec}\gtrsim2\frac{\mathcal L^{2}}{\epsilon T}
\end{eqnarray} for the continuously monitored and absorbing-potential implementations.
This is the measurement-based counterpart of the resource estimation for coherent CD, where the auxiliary control obeys $\|A(t)|n(t)\rangle\|=\sqrt{g(t)}$ \cite{delCampo2012,Funo2017}, so that its time-integrated amplitude is bounded by the same length $\mathcal L$. The two routes are thus comparable in geometric cost, and the practical choice is dictated by which resource (a fast monitoring channel, or a engineered, generically nonlocal, many-body control term) is available on a given platform. In particular, the Zeno route is not free from the geometric cost of a shortcut, which is fundamental and dictated by information  geometry. Rather, it converts it from coherent control amplitude into measurement bandwidth.

\newpage

\widetext
\newpage

\begin{center}
\text{\bf\large Supplemental Material}
\end{center}
%setcounter{tocdepth}{1}
%\tableofcontents

In this Supplemental Material, we use  equation labels prefixed by ``S'', keeping unprefixed labels for equations in the main text. 
Section~\ref{sec:CD} presents the correspondence between the stroboscopic protocol and counterdiabatic (CD) driving, including the derivation of the factor $\tfrac12$ in the geometric term of the Zeno generator for a general projection-valued measure (PVM). Section~\ref{sec:LZ} works through a two-level Landau--Zener-like protocol, explicitly showing that the stroboscopic, continuous-monitoring, and complex-absorbing-potential (CAP) routes all yield the same effective generator within the moving Zeno subspace, distinct from the coherent CD evolution. Section~\ref{sec:multi-rank} treats a three-level system in which the protected subspace is multi-rank and non-spectral, demonstrating Kato--Avron transport beyond what standard CD can generate. Section~\ref{sec:berry} presents an application of immediate practical interest: the realization of Berry-phase gates of arbitrary geometric content at finite speed via stroboscopic Zeno monitoring of the instantaneous ground state, without  Hamiltonian engineering of the CD control field. Section~\ref{sec:leakage} provides a step-by-step derivation of the leakage operators governing the departure from the ideal Zeno limit, of their equality, and of the resulting generalized Schulman relation. Section~\ref{sec:SME} derives the co-moving form of the stochastic master equation (SME) used in the main text, the inter-sector decoherence rates, and the leakage operator that governs the leading departure from the Zeno limit under continuous monitoring. Section~\ref{sec:PVM} describes the operational implementation of nonselective stroboscopic PVM measurements on a single quantum system, and Sec.~\ref{sec:multiCAP} generalizes the CAP  construction to an arbitrary number of monitored sectors. 

The examples and detailed derivations that follow illustrate complementary aspects of the framework introduced in the main text. 
Taken together, they make evident that the Zeno realizations of shortcuts to adiabaticity are not merely alternative implementations of the conventional CD based on Hamiltonian engineering. They are a strictly broader family of measurement- and dissipation-based control protocols, with regimes of advantage that are physically and operationally distinct from those of coherent CD.

\section{Operational implementation of nonselective stroboscopic PVM measurements}\label{sec:PVM}

In the stroboscopic construction, we repeatedly apply the completely positive map
\begin{equation}
\rho \mapsto \sum_n P_n(t_k)\,\rho\, P_n(t_k),
\label{eq:nonselective-map}
\end{equation}
corresponding to a full PVM $\{P_n(t_k)\}$.
This map does not require averaging over different experimental runs.
It can be implemented on a single quantum system by coupling the system to an ancillary probe (Stinespring dilation),
measuring the probe, and resetting it between steps.

Let the probe Hilbert space admit an orthonormal pointer basis $\{|n\rangle_A\}$.
Prepare the probe in a reference state $|0\rangle_A$ and apply a unitary interaction of the form
\begin{equation}
U_k = \sum_n P_n(t_k)\otimes W_n,
\qquad
W_n|0\rangle_A = |n\rangle_A,
\label{eq:stinespring-unitary}
\end{equation}
which entangles the system sectors with orthogonal probe states.
Acting on an arbitrary system state $\rho$, this unitary produces
\begin{equation}
U_k \big(\rho\otimes |0\rangle\langle 0|\big) U_k^\dagger
=
\sum_{n,m} P_n(t_k)\,\rho\,P_m(t_k)
\otimes |n\rangle\langle m|.
\label{eq:entangled-state}
\end{equation}
A projective measurement of the probe in the pointer basis $\{|n\rangle_A\}$ yields a definite outcome $n_k$
and collapses the system to $P_{n_k}(t_k)\rho P_{n_k}(t_k)/p_{n_k}$.
The probe is then reset to $|0\rangle_A$, and the evolution proceeds to the next stroboscopic interval.
However, if the measurement outcome is not used to condition subsequent control operations (nonselective protocol),
the reduced state of the system after the probe measurement is obtained by tracing over the probe and results in Eq.~\eqref{eq:nonselective-map}.
This implies discarding the classical measurement record while maintaining the physical system.
No restart of the experiment is required, and the system continues to evolve within whichever sector
was realized in that particular run.

\section{Relation between the stroboscopic protocol and counterdiabatic driving}\label{sec:CD}

\subsection{The geometric term of the Zeno generator for a general PVM}

Let $\{P_n(t)\}$ be a complete set of orthogonal differentiable projectors, satisfying
$P_nP_m=\delta_{nm}P_n$ and $\sum_nP_n=\mathbb I$. Upon repeating the stroboscopic
expansion of the main text for the PVM propagator
\begin{equation}
U_N(T)=\prod_k\sum_nP_n(t_{k+1})e^{-iH(t_k)\delta t}P_n(t_k), 
\end{equation}
and using
$P_n\dot P_nP_n=0$, each step contributes the geometric operator
$S\equiv\sum_n\dot P_n(t)P_n(t)$.
Differentiating $\sum_nP_n=\mathbb I$ gives $\sum_n\dot P_n=0$. Further, 
differentiating $P_n^2=P_n$ gives $\dot P_n=\dot P_nP_n+P_n\dot P_n$. Combining
the two,
\begin{equation}
S+S^\dagger=\sum_n\bigl(\dot P_nP_n+P_n\dot P_n\bigr)=\sum_n\dot P_n=0 ,
\end{equation}
so $S$ is anti-Hermitian. As a result, 
$\sum_n[\dot P_n,P_n]=\sum_n(\dot P_nP_n-P_n\dot P_n)=S-S^\dagger=2S$. Hence, the
geometric term admits the two equivalent forms
\begin{equation}
A=i\sum_n \dot P_n(t)\,P_n(t)
=\tfrac{i}{2}\sum_n\bigl[\dot P_n(t),P_n(t)\bigr],
\label{eq:A-two-forms}
\end{equation}
which is the origin of the factor $1/2$ in Eq.~(5) of the main text. For
the two-outcome PVM $\{P,Q\}$ one has $[\dot Q,Q]=[\dot P,P]$, so that
$A=i[\dot P,P]$ and Eq.~(5) reduces to Eq.~(4).

\subsection{Reduction to the counterdiabatic Hamiltonian}

Consider now the PVM built from the
instantaneous eigenprojectors of the system Hamiltonian. Let
\begin{equation}
H(t)=\sum_n E_n(t)\,P_n(t),\qquad P_n(t)=\ket{n(t)}\bra{n(t)},
\end{equation}
with a nondegenerate spectrum and orthogonal, differentiable, unit-rank
projectors.
From the general result, the effective Zeno Hamiltonian is
$H_{\rm Z}=\sum_nP_nHP_n+A$. The dynamical term collapses onto the
spectral decomposition,
\begin{equation}
\sum_n P_n H P_n=\sum_n E_n(t)\,P_n(t)=H(t),
\end{equation}
so the eigenbasis diagonalizes the projected dynamics with no dynamical leakage.
Differentiating $P_n=\ket{n}\bra{n}$ yields
$\dot P_n=\ket{\dot n}\bra{n}+\ket{n}\bra{\dot n}$, and, using
$\braket{\dot n}{n}=-\braket{n}{\dot n}$, it follows that
\begin{equation}
\dot P_n P_n=\ket{\dot n}\bra{n}-\braket{n}{\dot n}\,\ket{n}\bra{n}.
\label{eq:PdotP}
\end{equation}
Substituting \eqref{eq:PdotP} into \eqref{eq:A-two-forms} gives the
auxiliary term in closed form,
\begin{equation}
A=i\sum_n\Bigl(\ket{\dot n}\bra{n}-\braket{n}{\dot n}\,\ket{n}\bra{n}\Bigr),
\label{eq:A-Berry}
\end{equation}
so that
\begin{equation}
H_{\rm Z}=H(t)+i\sum_n\Bigl(\ket{\dot n}\bra{n}
-\braket{n}{\dot n}\,\ket{n}\bra{n}\Bigr).
\label{eq:HZ-CD}
\end{equation}
Equation~\eqref{eq:A-Berry} is precisely the counterdiabatic
term (for transitionless quantum driving), also known as the adiabatic gauge
potential, derived by Demirplak and Rice and independently by Berry,
including the diagonal Berry-connection contribution
$-\braket{n}{\dot n}\ket{n}\bra{n}$ \cite{Demirplak2003,Demirplak2005,Demirplak2008,Berry2009}. Thus, $H_{\rm Z}=H+H_{\rm CD}$
drives the system exactly along the instantaneous eigenstates of $H(t)$
without requiring slow driving.

However, note that the channel implemented by the nonselective Zeno
protocol differs from the unitary generated by $H_{\rm Z}$. For an initial pure
state $|\psi(0)\rangle=\sum_nc_n(0)|n(0)\rangle$, CD produces the coherent
superposition
\begin{equation}
|\psi_{\rm CD}(t)\rangle=\sum_nc_n(0)\,
e^{-i\int_0^tE_n(s)ds-\int_0^t\braket{n(s)}{\dot n(s)}ds}\,|n(t)\rangle ,
\end{equation}
whereas the adaptive Zeno protocol produces the mixed state
$\rho_{\rm Z}(t)=\sum_n|c_n(0)|^2P_n(t)$. The two agree on the populations
$\{|c_n(0)|^2\}$ and differ only on the inter-sector coherences, which the
nonselective measurement destroys. In particular, they implement the same map
whenever $\rho_0$ is block diagonal in $\{P_n(0)\}$.

\section{Two-level avoided crossing: equivalence of the three Zeno routes}\label{sec:LZ}

Consider a qubit driven through an avoided crossing,
\begin{equation}
H(t)=\tfrac{1}{2}\Delta(t)\,\sigma_z+\tfrac{1}{2}\Omega\,\sigma_x,
\label{eq:LZ-H}
\end{equation}
with $\Delta(t)$ swept and $\Omega>0$ fixed. The instantaneous eigenstates
\begin{equation}
|0(t)\rangle=\cos[\tfrac{\theta(t)}{2}]\,|\!\!\uparrow\rangle+\sin[\tfrac{\theta(t)}{2}]\,|\!\!\downarrow\rangle,
\qquad
|1(t)\rangle=-\sin[\tfrac{\theta(t)}{2}]\,|\!\!\uparrow\rangle+\cos[\tfrac{\theta(t)}{2}]\,|\!\!\downarrow\rangle,
\label{eq:adiabatic-basis}
\end{equation}
are parametrized by $\tan\theta(t)=\Omega/\Delta(t)$, with instantaneous energies $E_{0,1}(t)=\mp\tfrac{1}{2}\sqrt{\Delta(t)^2+\Omega^2}$. The eigenprojectors are $P_n(t)=|n(t)\rangle\langle n(t)|$, and a direct computation yields
\begin{equation}
\dot P_0=\frac{\dot\theta}{2}\,(|1\rangle\langle 0|+|0\rangle\langle 1|),
\qquad
[\dot P_0,P_0]=\frac{\dot\theta}{2}\,(|1\rangle\langle 0|-|0\rangle\langle 1|),
\label{eq:dotP-LZ}
\end{equation}
with $\dot P_1=-\dot P_0$ (from $P_0+P_1=\mathbb{I}$) and $[\dot P_1,P_1]=[\dot P_0,P_0]$.

In the conventional CD approach \cite{Demirplak2003,Demirplak2005,Demirplak2008,Berry2009}
the CD Hamiltonian for this system is \cite{Torrontegui2013}
\begin{equation}
H_{\rm CD}(t)=H(t)+\tfrac{1}{2}\dot\theta(t)\,\sigma_y,
\label{eq:HCD-LZ}
\end{equation}
which generates exact transitionless evolution along the adiabatic states (\ref{eq:adiabatic-basis}), with both dynamical and geometric phases.

In the stroboscopic Zeno route, at each time $t_k=k\,\delta t$, we perform a nonselective PVM $\{P_0(t_k),P_1(t_k)\}$. By Eq.~(5) of the main text, the effective Zeno generator is
\begin{equation}
H_{\rm Z}(t)=\sum_n P_n H P_n+\tfrac{i}{2}\sum_n[\dot P_n,P_n]
=H(t)+\tfrac{1}{2}\dot\theta(t)\,\sigma_y,
\label{eq:HZ-LZ}
\end{equation}
which coincides with $H_{\rm CD}$ in (\ref{eq:HCD-LZ}).  However, the global evolution is not the unitary generated by $H_{\rm CD}$.  The nonselective measurements implement a pinching map $\Lambda(\rho)=\sum_n P_n(0)\rho P_n(0)$ at the initial time, followed by coherent evolution within each sector. For a pure initial state $|\psi(0)\rangle=c_0|0(0)\rangle+c_1|1(0)\rangle$ the state at time $T$ is
\begin{equation}
\rho_{\rm Z}(T)=|c_0|^2 P_0(T)+|c_1|^2 P_1(T),
\label{eq:rho-Z-LZ}
\end{equation}
while CD gives $|\psi_{\rm CD}(T)\rangle\langle\psi_{\rm CD}(T)|$, a pure state with all coherences in the adiabatic basis preserved. The two trajectories agree on the diagonal populations $\{|c_n|^2\}$ but differ in the off-diagonal coherences $c_n c_m^*$, which the adaptive Zeno measurements remove.

An alternative approach relies on continuous measurements. 
Consider the continuous monitoring of the observable $X(t)=P_1(t)-P_0(t)$. The conditioned state $\rho_c$ obeys the SME [Eq.~(10) of the main text] with the Hermiticity-preserving innovation term $\mathcal{H}[X(t)]\rho=\{X(t),\rho\}-2\langle X(t)\rangle\rho$. In the co-moving frame defined by the intertwiner $W(t)$ with $\dot W=K W$ and $K(t)=\tfrac{1}{2}\sum_n[\dot P_n,P_n]=\tfrac{1}{2}\dot\theta\,(|1\rangle\langle 0|-|0\rangle\langle 1|)$, the projectors become time independent and the dissipator $\kappa\mathcal D[X_0]$ dephases inter-sector coherences at the rate $\gamma_{\rm dec}=\tfrac{\kappa}{2}(x_1-x_0)^2=2\kappa$, cf.\ Sec.~\ref{sec:SME}. The remaining intra-sector dynamics is generated, in the laboratory frame, by Eq.~(\ref{eq:HZ-LZ}). The strong-measurement limit, with $\kappa\to\infty$, therefore agrees with the stroboscopic result, and the correspondence of Eq.~(14) of the main text gives $\delta t=2/\gamma_{\rm dec}=1/\kappa$ for this choice of eigenvalues.

Finally, let us consider the STA engineering by a CAP. 
We choose $P(t)\equiv P_0(t)$ as the protected subspace and $Q(t)=P_1(t)$ as the absorbing one. The non-Hermitian Hamiltonian is
\begin{equation}
H_{\rm nh}(t)=H(t)-i\kappa\,P_1(t).
\label{eq:Hnh-LZ}
\end{equation}
At leading order in $1/\kappa$, the adiabatic elimination of the absorbing sector (see Sec.~\ref{sec:leakage}) yields the Kato--Avron Hamiltonian $H_{\rm eff}(t)=P_0 H P_0+i[\dot P_0,P_0]$ within $\mathrm{Ran}\,P_0(t)$. In the adiabatic basis,
\begin{equation}
P_0 H(t) P_0=E_0(t)\,P_0(t),
\qquad
i[\dot P_0,P_0]=\tfrac{i}{2}\dot\theta\,(|1\rangle\langle 0|-|0\rangle\langle 1|),
\end{equation}
which (acting on $\mathrm{Ran}\,P_0$) is precisely the part of $\tfrac{1}{2}\dot\theta\,\sigma_y$ that takes vectors in $\mathrm{Ran}\,P_0(t)$ along $\mathrm{Ran}\,P_0(t+dt)$. The resulting evolution within the protected subspace is identical to that produced by the stroboscopic and continuous-monitoring routes restricted to $\mathrm{Ran}\,P_0(t)$.

Away form the strict Zeno limit, leakage generally occurs. 
Because $P_0(t)$ is a spectral projector of $H(t)$, one has $QHP=0$ and the leakage operator of Eq.~(7) of the main text is purely geometric,
\begin{equation}
\Gamma=P_0\dot P_0\dot P_0P_0=\tfrac{1}{4}\dot\theta^2\,P_0 ,
\end{equation}
so that $p_{\rm leak}=\tfrac14\delta t^2\dot\theta^2$ per stroboscopic step and, by the generalized Schulman relation, $\tfrac{1}{2}\dot\theta^{2}/\gamma_{\rm dec}$ per unit time in the continuously monitored and CAP realizations. In all three routes, the departure from ideal Zeno transport is controlled by the same angular velocity $\dot\theta$ of the adiabatic basis.

In summary, all three Zeno routes produce the same effective generator within each sector and are characterized by the same leading correction away from the Zeno limit. They differ from the coherent CD evolution only in the suppression of inter-sector coherences. This can prove advantageous over coherent CD, as shown in the applications considered in the next two sections.

\section{Beyond CD: Kato--Avron transport in a non-spectral multi-rank subspace}\label{sec:multi-rank}

The Kato--Avron Hamiltonian acts on any moving subspace $\mathrm{Ran}\,P(t)$ with $\dot P\neq 0$. When $P(t)$ is not a spectral projector of $H(t)$, the resulting generator generalizes the standard CD, which is defined only for spectral projectors. This section provides a concrete example with a three-level example, as an illustration.

Consider a three-level system with $H_0=\mathrm{diag}(\epsilon,\epsilon,0)$, in the basis $\{|1\rangle,|2\rangle,|3\rangle\}$. The two-dimensional ``code'' subspace $\mathrm{span}\{|1\rangle,|2\rangle\}$ is degenerate. Define a one-parameter family of two-dimensional projectors
\begin{equation}
P(t)=V(t)\,P_{\rm code}\,V^\dagger(t),
\qquad
P_{\rm code}=|1\rangle\langle 1|+|2\rangle\langle 2|,
\label{eq:Pt-3level}
\end{equation}
where $V(t)$ is a unitary that rotates the code subspace into the full three-dimensional Hilbert space along a chosen path. For concreteness, take
\begin{equation}
V(t)=\exp\!\left[-i\,\phi(t)\,\bigl(|3\rangle\langle 2|+|2\rangle\langle 3|\bigr)\right]\,
\exp\!\left[-i\,\xi(t)\,\bigl(|2\rangle\langle 1|+|1\rangle\langle 2|\bigr)\right],
\label{eq:Vt}
\end{equation}
parametrized by two angles $\phi(t),\xi(t)$. The instantaneous code subspace $\mathrm{Ran}\,P(t)$ is, in general, not a sum of eigenspaces of $H_0$ (or any time-independent reference Hamiltonian), since $V(t)$ mixes the degenerate manifold with the $|3\rangle$ sector.

We first notice the failure of the standard CD in this context.
Since $P(t)$ is not a spectral projector of $H_0$, the conventional CD recipe $H_{\rm CD}=H_0+\frac{i}{2}\sum_n[\dot P_n,P_n]$ applied to the eigenprojectors of $H_0$ will not produce transport along the chosen $P(t)$. There is no standard CD Hamiltonian for this protocol.

Yet, Kato--Avron transport can be implemented via adaptive quantum Zeno measurements.
The Zeno protocol monitors $P(t)$, with the effective generator within the protected subspace
\begin{equation}
H_{\rm Z}(t)=P(t)\,H_0\,P(t)+i[\dot P(t),P(t)].
\label{eq:HZ-multi-rank}
\end{equation}
The first term is the projection of the physical Hamiltonian onto the moving code subspace and the second term is the geometric Kato--Avron connection. Computing $\dot P$ from (\ref{eq:Pt-3level}),
\begin{equation}
\dot P=[\dot V V^\dagger,\,P],
\qquad
\dot V V^\dagger=-i\dot\xi(t)(|2\rangle\langle 1|+|1\rangle\langle 2|)
-i\dot\phi(t)\,V_\xi\,(|3\rangle\langle 2|+|2\rangle\langle 3|)\,V_\xi^\dagger,
\label{eq:dotV}
\end{equation}
with $V_\xi=\exp[-i\xi(|2\rangle\langle 1|+|1\rangle\langle 2|)]$. The intertwiner relation $[K,P]=\dot P$ is satisfied by $K(t)=\dot V V^\dagger$, anti-Hermitian by construction.

The dynamics within $\mathrm{Ran}\,P(t)$ is a non-Abelian holonomy. A closed loop $\phi(0)=\phi(T)$, $\xi(0)=\xi(T)$ accumulates a unitary
\begin{equation}
U_{\rm hol}=\mathcal{T}\exp\!\left[-i\int_0^T H_{\rm Z}(t)\,dt\right]\bigg|_{\mathrm{Ran}\,P_{\rm code}},
\end{equation}
which is, in general, non-trivial and depends on the loop, not on its parametrization. This is precisely the dynamical realization of a non-Abelian holonomy, generated by the Kato--Avron Hamiltonian, with no coherent CD analog. In the adiabatic limit, or under the Zeno protocol that simulates it via measurements, $U_{\rm hol}$ acts as a unitary gate on the code subspace, of direct relevance to holonomic quantum computation \cite{Zanardi1999,Burgarth2013}.

Here, in contrast with the situation of Sec.~\ref{sec:LZ}, the projector is not spectral, $QH_0P\neq0$, and the leakage operator of Eq.~(7) of the main text retains all three contributions,
\begin{equation}
\Gamma=PH_0QH_0P+P\dot P\dot PP-i\bigl(PH_0Q\dot PP-P\dot PQH_0P\bigr),
\end{equation}
so that the error budget of the resulting holonomic gate contains a dynamical piece set by the energy splitting $\epsilon$ in addition to the geometric piece set by the speed of the loop.

In conclusion, for any task that requires controlled transport within a non-spectral multi-rank subspace (such as holonomic gates in degenerate manifolds, controlled rotations within a code space of an error-correcting code, and transport along a sub-bundle in parameter space), the Kato--Avron Hamiltonian provides the generator, and the three Zeno routes provide its physical realizations. By contrast, conventional CD does not. Thus, there are scenarios where the engineering of STA by Zeno measurements is preferable to the conventional case based on Hamiltonian engineering. 

\section{Application: accelerated Berry-phase gates via Zeno protection}\label{sec:berry}

A direct and practically important application of the framework is the dynamical realization of geometric (Berry) phase gates on a closed adiabatic loop, at finite speed and without synthesis of the counterdiabatic control field \cite{Burgarth2013}.

Consider a two-level system with
\begin{equation}
H(t)=\tfrac{\Omega}{2}\,\hat{\bm{n}}(t)\!\cdot\!\bm{\sigma},
\qquad
\hat{\bm{n}}(t)=\bigl(\sin\theta(t)\cos\phi(t),\,\sin\theta(t)\sin\phi(t),\,\cos\theta(t)\bigr),
\label{eq:H-Berry}
\end{equation}
with constant $\Omega>0$. The polar angles $(\theta(t),\phi(t))$ trace a closed loop $\mathcal{C}$ on the Bloch sphere, $\hat{\bm{n}}(T)=\hat{\bm{n}}(0)$. The instantaneous ground state $|g(t)\rangle$ and the corresponding rank-one projector $P_g(t)=|g(t)\rangle\langle g(t)|$ inherit the same periodicity. In the slow adiabatic limit, the ground state acquires a dynamical phase together with the Berry phase
\begin{equation}
\gamma_{\rm B}(\mathcal{C})\,=\,-\tfrac{1}{2}\oint_\mathcal{C}(1-\cos\theta)\,d\phi\,=\,-\tfrac{1}{2}\,\Omega_{\rm sol}(\mathcal{C}),
\label{eq:Berry-formula}
\end{equation}
where $\Omega_{\rm sol}(\mathcal{C})$ is the solid angle enclosed by $\mathcal{C}$.

In the conventional CD realization,  the counterdiabatic Hamiltonian for (\ref{eq:H-Berry}) is
\begin{equation}
H_{\rm CD}(t)=H(t)+\tfrac{1}{2}\,\bigl(\dot\theta\,\hat{\bm{e}}_{\phi}-\dot\phi\sin\theta\,\hat{\bm{e}}_{\theta}\bigr)\!\cdot\!\bm{\sigma},
\label{eq:HCD-Berry}
\end{equation}
where $(\hat{\bm{e}}_\theta,\hat{\bm{e}}_\phi)$ are the spherical unit vectors associated with $\hat{\bm{n}}$. The CD control is a vector field tangent to the Bloch sphere along $\mathcal{C}$, with strength proportional to the angular velocity of $\hat{\bm{n}}(t)$. Implementing it requires phase-coherent use of two simultaneous control fields whose orientations and amplitudes vary along the loop. In platforms where the natural control is a single rotating field (e.g., NV centers, trapped ions, spin qubits with a vector magnetic field), this requires non-trivial auxiliary hardware.

The Zeno realization provides an alternative.
The Zeno protocol monitors $P_g(t)$ stroboscopically, or, equivalently, performs continuous monitoring of $\hat{\bm{n}}(t)\!\cdot\!\bm{\sigma}$ in the strong-measurement limit. The effective generator within the protected subspace is the Kato--Avron Hamiltonian [Eq.~(4) of the main text], $H_{\rm Z}(t)=P_g H P_g + i[\dot P_g,P_g]$, which coincides with $H_{\rm CD}(t)$ on $\mathrm{Ran}\,P_g(t)$. The final state, conditioned on staying in the protected subspace throughout the loop, is therefore
\begin{equation}
|\psi_{\rm Z}(T)\rangle = e^{-i\int_0^T E_g(s)\,ds\,+\,i\gamma_{\rm B}(\mathcal{C})}\,|g(0)\rangle,
\label{eq:psiZ-Berry}
\end{equation}
acquiring the same Berry phase (\ref{eq:Berry-formula}) as the adiabatic limit,  at any total time $T$. The key practical point is that the Zeno protocol requires only a single observable per time step, i.e., the spin component along $\hat{\bm{n}}(t)$. There is no need for an auxiliary control field along the tangent directions $\hat{\bm{e}}_{\theta,\phi}$.

By way of example, for a polar-cap loop of half-angle $\alpha$ traced uniformly in time, $\theta(t)=\alpha$ and $\phi(t)=2\pi t/T$, the enclosed solid angle is $\Omega_{\rm sol}=2\pi(1-\cos\alpha)$, giving
\begin{equation}
\gamma_{\rm B}=-\pi(1-\cos\alpha).
\label{eq:Berry-cap}
\end{equation}
For $\alpha=\pi/2$ (equatorial loop), $\gamma_{\rm B}=-\pi$, implementing a logical $Z$ gate on the ground state up to dynamical phase. Conversely, for $\alpha=\pi/3$, $\gamma_{\rm B}=-\pi/2$, realizing an $S$ gate. The geometric phase depends only on $\mathcal{C}$, and not on the speed, as small fluctuations of $\dot\phi(t)$ that preserve the loop integral leave $\gamma_{\rm B}$ invariant.

\paragraph*{Resource cost and leakage.---}
The Zeno protocol is governed by the stroboscopic interval $\delta t$ or, equivalently, by the inter-sector decoherence rate $\gamma_{\rm dec}$ in the continuous and CAP formulations. Since $P_g(t)$ is a spectral projector of $H(t)$, the dynamical contribution vanishes identically, $Q_gHP_g=0$, and the leakage operator of Eq.~(7) of the main text is purely geometric,
\begin{equation}
\Gamma=P_g\dot P_g\dot P_g P_g=\tfrac{1}{4}\bigl(\dot\theta^2+\dot\phi^2\sin^2\theta\bigr)\,P_g,
\qquad Q_g=\mathbb{I}-P_g .
\label{eq:Gamma-Berry}
\end{equation}
The right-hand side is precisely the Fubini--Study metric $g(t)$ of the path traced by $|g(t)\rangle$, so that the  leakage probability per step is
\begin{equation}
p_{\rm leak}\,=\,\delta t^2\,g(t),
\qquad
g(t)=\tfrac{1}{4}\bigl(\dot\theta^2+\dot\phi^2\sin^2\theta\bigr),
\end{equation}
which is suppressed by taking $\delta t$ small compared with $1/|\dot{\hat{\bm{n}}}|$. Note that the leakage is controlled by the speed of the loop alone and not by $\Omega$. The energy scale of the Hamiltonian drops out because the monitored projector is spectral.

For the polar-cap loop above, $g=\pi^2\sin^2\alpha/T^2$ is constant, and summing over $N=T/\delta t$ steps gives the total leakage
\begin{equation}
p^{\rm tot}_{\rm leak}=\delta t\int_0^Tg\,dt=\frac{\pi^2\sin^2\alpha}{N}=\frac{\mathcal L^2}{N},
\end{equation}
where $\mathcal L=\int_0^T\sqrt{g}\,dt=\pi\sin\alpha$ is the Fubini--Study length of the loop. This saturates the geometric bound derived in the End Matter, as it must for constant-speed dragging. Demanding $p^{\rm tot}_{\rm leak}\le\epsilon$ therefore requires $N\ge\pi^2\sin^2\alpha/\epsilon$ measurements, regardless the gate duration $T$. Due to the generalized Schulman relation $\delta t=2/\gamma_{\rm dec}$, the same accuracy is obtained in the continuously monitored and CAP implementations for $\gamma_{\rm dec}\ge2\pi^2\sin^2\alpha/(\epsilon T)$.

The Zeno protocol thus realizes Berry-phase gates of arbitrary geometric content at arbitrary speed, without relying on CD control fields, requiring only a stroboscopic measurement of a single spin direction per step. This is the operational advantage in platforms where projective single-spin measurements along an arbitrary direction are routine (e.g., NV centers under optically detected magnetic resonance, trapped-ion fluorescence imaging), but using coherent control fields with time-dependent amplitudes is not. Generalizations to non-Abelian holonomies follow directly by replacing the rank-one projector $P_g(t)$ with a multi-rank projector onto a degenerate manifold, as in Sec.~\ref{sec:multi-rank}.

\section{Derivation of the leakage operators and of the generalized Schulman relation}\label{sec:leakage}

This section analyzes in detail the leakage operators that govern the
leading departure from the ideal Zeno limit in two of the protocols of the
main text: the stroboscopic protocol at finite measurement interval $\delta t$,
and the CAP protocol at a finite absorption
strength $\kappa$. We show that the two operators are not merely analogous but
identical, and that matching the corresponding generators promotes the
Schulman correspondence to the model-independent relation
$\delta t=2/\gamma_{\rm dec}$.

\subsection{Useful projector identities}

Differentiating $P^2=P$ gives $\dot P P+P\dot P=\dot P$. Multiplying on both
sides by $P$ and using $P^2=P$ yields $2P\dot P P=P\dot P P$, hence \cite{Kato1950}
\begin{equation}
P\dot P P=0,
\label{eq:PdPP}
\end{equation}
which states that $\dot P$ has no component within the subspace. Two immediate
corollaries express that $\dot P$ is purely off-diagonal,
\begin{equation}
\dot P P=(P+Q)\dot P P=Q\dot P P,
\qquad
P\dot P=P\dot P(P+Q)=P\dot P Q,
\label{eq:offdiag}
\end{equation}
where \eqref{eq:PdPP} was used. From de definition of $A$ and \eqref{eq:PdPP}, we also
record the three contractions of $A$ below,
\begin{equation}
AP=i\dot P P,
\qquad
PA=-iP\dot P,
\qquad
PAP=iP[\dot P,P]P=0.
\label{eq:Acontract}
\end{equation}

\subsection{Stroboscopic leakage operator}\label{sec:Gdt}

A single conditioned step from $t$ to $t+\delta t$ (project, evolve, project) is described by 
the non-unitary map
\begin{equation}
\Omega(t)=P(t+\delta t)\,e^{-iH\delta t}\,P(t),
\label{eq:Omega}
\end{equation}
where we approximate $H(t')\simeq H(t)$ on the short interval. Expanding both
factors to second order, with all quantities on the right evaluated at $t$, one finds 
\begin{equation}
P(t+\delta t)=P+\dot P\,\delta t+\tfrac12\ddot P\,\delta t^2+\mathcal O(\delta t^3),
\qquad
e^{-iH\delta t}=\mathbb I-iH\delta t-\tfrac12 H^2\delta t^2+\mathcal O(\delta t^3).
\end{equation}
It thus follows that the map is given by
\begin{equation}
\Omega=P+\delta t\,\omega_1+\delta t^2\,\omega_2+\mathcal O(\delta t^3),
\qquad
\omega_1=\dot P P-iPHP,
\qquad
\omega_2=-\tfrac12 PH^2P-i\dot P HP+\tfrac12\ddot P P.
\label{eq:Omega-exp}
\end{equation}
Note that even in the Zeno subspace, $\Omega(t)$ is not unitary for finite $\delta t$, and its norm loss
encodes the probability of leaving the monitored subspace between measurements.

Since every term in \eqref{eq:Omega-exp} ends in $P$, one has $\Omega=\Omega P$,
so $\Omega^\dagger\Omega=P\Omega^\dagger\Omega P$ is automatically restricted to
the subspace. We analyze it order by order.

To order $\delta t$, the contribution is $P\omega_1 P+P\omega_1^\dagger P$. Using $P\dot P P=0$,
\begin{equation}
P\omega_1 P=-iPHP,
\qquad
P\omega_1^\dagger P=+iPHP
\quad\Longrightarrow\quad
P\omega_1 P+P\omega_1^\dagger P=0,
\end{equation}
confirming that the survival probability deviates from unity only at
$\mathcal O(\delta t^2)$.

To order $\delta t^2$, the contribution is $P\omega_1^\dagger\omega_1 P+P\omega_2 P+P\omega_2^\dagger P$.
For the first piece, expanding and discarding every term containing
$P\dot P P=0$,
\begin{equation}
P\omega_1^\dagger\omega_1 P
=P(P\dot P+iPHP)(\dot P P-iPHP)P
=P\dot P\dot P P+PHPHP.
\end{equation}
For the remaining pieces,
\begin{equation}
P\omega_2 P+P\omega_2^\dagger P
=-PH^2P+P\ddot P P-iP\dot P HP+iPH\dot P P,
\end{equation}
where the antisymmetric $\dot P H$ terms survive because they are not protected
by $P\dot P P=0$. To simplify the sum, we insert $\mathbb I=P+Q$, so that
\begin{equation}
PHPHP-PH^2P=PHPHP-PH(P+Q)HP=-PHQHP.
\label{eq:dyn-reduce}
\end{equation}
Differentiating \eqref{eq:PdPP} twice and sandwiching with $P$ gives
\begin{equation}
P\ddot P P=-2P\dot P\dot P P.
\label{eq:PddotP}
\end{equation}
Collecting all $\mathcal O(\delta t^2)$ terms with \eqref{eq:dyn-reduce} and
\eqref{eq:PddotP}, and using the off-diagonal identities \eqref{eq:offdiag} to
insert $Q$ via $PH\dot P P=PHQ\dot P P$ and $P\dot P HP=P\dot P QHP$,
\begin{equation}
P\Omega^\dagger\Omega P
=P-\delta t^2\,\Gamma_{\delta t}+\mathcal O(\delta t^3),
\qquad
\Gamma_{\delta t}
=PHQHP+P\dot P\dot P P
-i\bigl(PHQ\dot P P-P\dot P QHP\bigr).
\label{eq:Gdt}
\end{equation}
For a normalized state $|\psi\rangle\in\operatorname{Ran}P$, the 
leakage probability per step
$p_{\rm leak}=1-\langle\psi|P\Omega^\dagger\Omega P|\psi\rangle$ is therefore
\begin{equation}
p_{\rm leak}=\delta t^2\,\langle\psi|\Gamma_{\delta t}|\psi\rangle
+\mathcal O(\delta t^3),
\label{eq:pleak-SM}
\end{equation}
in agreement with Eq.~(20) of the End Matter.

The leakage operator admits a manifestly positive form.
Expanding $P(H-A)Q(H-A)P$ with the contractions \eqref{eq:Acontract},
\begin{equation}
PAQAP=(-iP\dot P)\,Q\,(i\dot P P)=P\dot P Q\dot P P=P\dot P\dot P P,
\qquad
PHQAP=iPHQ\dot P P,
\qquad
PAQHP=-iP\dot P QHP,
\end{equation}
where $P\dot P Q\dot P P=P\dot P\dot P P$ follows from
$Q=\mathbb I-P$ and \eqref{eq:PdPP}. Hence,
\begin{equation}
P(H-A)Q(H-A)P
=PHQHP+P\dot P\dot P P-i\bigl(PHQ\dot P P-P\dot P QHP\bigr),
\end{equation}
which matches \eqref{eq:Gdt} term by term. Therefore,
\begin{equation}
\Gamma_{\delta t}=P\bigl(H-i[\dot P,P]\bigr)Q\bigl(H-i[\dot P,P]\bigr)P
=\bigl[Q(H-A)P\bigr]^\dagger\bigl[Q(H-A)P\bigr]\;\ge0 .
\label{eq:Gdt-square}
\end{equation}

\subsection{Absorbing-potential leakage operator}\label{sec:Gk}

Consider now the time evolution under the non-Hermitian Hamiltonian
\begin{equation}
H_{\rm nh}(t)=H(t)-i\kappa\,Q(t),\qquad \kappa>0.
\label{eq:Hnh-SM}
\end{equation}
We are interested in the effective generator within $\operatorname{Ran}P$ to leading order
in $1/\kappa$.

Let us introduce the intertwiner $W(t)$ defined by $\dot W=KW$, $W(0)=\mathbb I$, with
the anti-Hermitian generator $K=[\dot P,P]$, which satisfies $[K,P]=\dot P$. The intertwiner $W(t)$
transports the projectors, $P(t)=W P_0 W^\dagger$, $Q(t)=W Q_0 W^\dagger$, where
$P_0=P(0)$ is time independent. In the co-moving frame
$|\tilde\psi\rangle=W^\dagger|\psi\rangle$, substituting $|\psi\rangle=W|\tilde\psi\rangle$
into $i\partial_t|\psi\rangle=H_{\rm nh}|\psi\rangle$ and using $W^\dagger Q W=Q_0$,
\begin{equation}
i\partial_t|\tilde\psi\rangle=\bigl(\tilde H-i\kappa Q_0\bigr)|\tilde\psi\rangle,
\qquad
\tilde H=W^\dagger H W-iW^\dagger\dot W,
\label{eq:tilde-eom}
\end{equation}
so that the absorbing projector $Q_0$ is now time independent.

We next proceed by adiabatic elimination and decompose $|\tilde\psi\rangle=|\tilde\psi_P\rangle+|\tilde\psi_Q\rangle$ with
$|\tilde\psi_P\rangle=P_0|\tilde\psi\rangle$ and
$|\tilde\psi_Q\rangle=Q_0|\tilde\psi\rangle$. Projecting \eqref{eq:tilde-eom}
onto $P_0$ and $Q_0$,
\begin{align}
i\dot{\tilde\psi}_P &=P_0\tilde H P_0\,\tilde\psi_P+P_0\tilde H Q_0\,\tilde\psi_Q,\\
i\dot{\tilde\psi}_Q &=Q_0\tilde H P_0\,\tilde\psi_P
+\bigl(Q_0\tilde H Q_0-i\kappa Q_0\bigr)\tilde\psi_Q.
\end{align}
For large $\kappa$ the absorbing sector is overdamped. Setting
$\dot{\tilde\psi}_Q\simeq0$ and inverting the $Q$-block,
$(Q_0\tilde H Q_0-i\kappa Q_0)^{-1}=\tfrac{i}{\kappa}Q_0+\mathcal O(\kappa^{-2})$, one finds
\begin{equation}
|\tilde\psi_Q\rangle\simeq\frac{1}{i\kappa}\,Q_0\tilde H P_0\,|\tilde\psi_P\rangle
+\mathcal O(\kappa^{-2}),
\end{equation}
which substituted back yields the co-moving effective generator
\begin{equation}
\tilde H_{\rm eff}=P_0\tilde H P_0
-\frac{i}{\kappa}\,P_0\tilde H Q_0\tilde H P_0+\mathcal O(\kappa^{-2}).
\label{eq:tilde-Heff}
\end{equation}

For any co-moving operator, the laboratory-frame image is $W(\cdot)W^\dagger$. In
particular, using $\dot W W^\dagger=K$,
\begin{equation}
W\tilde H W^\dagger
=W\bigl(W^\dagger H W-iW^\dagger\dot W\bigr)W^\dagger
=H-i\dot W W^\dagger
=H-iK
=H-A,
\label{eq:WHW}
\end{equation}
where the last step uses $iK=i[\dot P,P]=A$. Thus the off-diagonal coupling that
drives leakage carries the geometric term with a minus sign.

We next transform back to the laboratory frame.
From $i\partial_t|\tilde\psi_P\rangle=\tilde H_{\rm eff}|\tilde\psi_P\rangle$ and
$|\psi_P\rangle=W|\tilde\psi_P\rangle$, the laboratory-frame generator picks up the
frame term $i\dot W W^\dagger=iK$,
\begin{equation}
H_{\rm eff}=iK+W\tilde H_{\rm eff}W^\dagger.
\end{equation}
At leading order, with \eqref{eq:WHW} and $PAP=0$ from \eqref{eq:Acontract},
\begin{equation}
iK+W(P_0\tilde H P_0)W^\dagger=iK+P(H-A)P=PHP+iK=PHP+i[\dot P,P]=H_{\rm Z},
\end{equation}
so the leading geometric term is $+A$, as dictated by the frame term.
The $\mathcal O(1/\kappa)$ correction maps to
\begin{equation}
W\!\left(-\frac{i}{\kappa}P_0\tilde H Q_0\tilde H P_0\right)\!W^\dagger
=-\frac{i}{\kappa}\,P(H-A)\,Q\,(H-A)P .
\end{equation}
Writing $H_{\rm eff}=H_{\rm Z}-\tfrac{i}{\kappa}\Gamma_\kappa+\mathcal O(\kappa^{-2})$,
as in Eq.~(12) of the main text, identifies the leakage operator
\begin{equation}
\Gamma_\kappa=P\bigl(H-i[\dot P,P]\bigr)Q\bigl(H-i[\dot P,P]\bigr)P
=PHQHP+P\dot P\dot P P
-i\bigl(PHQ\dot P P-P\dot P QHP\bigr).
\label{eq:Gk}
\end{equation}

\subsection{Equality of the leakage operators and the generalized Schulman relation}

Comparing \eqref{eq:Gdt-square} with \eqref{eq:Gk}, the stroboscopic and
absorber leakage operators coincide exactly,
\begin{equation}
\Gamma_{\delta t}=\Gamma_\kappa\equiv\Gamma
=P\bigl(H-i[\dot P,P]\bigr)Q\bigl(H-i[\dot P,P]\bigr)P ,
\label{eq:equality}
\end{equation}
including their geometric cross terms. This identity holds at the level of
operators, with no terms neglected. The two square contributions represent,
respectively, dynamical leakage due to the couplings $QHP$ between the Zeno
subspace and its complement, and geometric leakage associated with the motion of
the projector. The cross terms are common to both routes and vanish when
$QHP=0$ (spectral projector) or $\dot P=0$ (static projector).

The two effective Hamiltonians read
\begin{equation}
H_{\rm eff}^{(\delta t)}=H_{\rm Z}-\frac{i\delta t}{2}\,\Gamma,
\qquad
H_{\rm eff}^{(\kappa)}=H_{\rm Z}-\frac{i}{\kappa}\,\Gamma ,
\label{eq:two-generators}
\end{equation}
so that matching the anti-Hermitian parts, $\tfrac{\delta t}{2}\Gamma=\tfrac{1}{\kappa}\Gamma$, requires
\begin{equation}
\delta t=\frac{2}{\kappa},
\label{eq:Schulman-SM}
\end{equation}
rather than the naive $\delta t=1/\kappa$. The prefactor in
$H_{\rm eff}^{(\delta t)}$ follows from \eqref{eq:pleak-SM}. Over $N=T/\delta t$
steps the norm squared decays at the rate $\delta t\langle\Gamma\rangle$, which
is precisely the rate produced by an anti-Hermitian generator
$-\tfrac{i\delta t}{2}\Gamma$.

The factor of two is easily verified on a two-level system with
$H=\Omega\sigma_x$, $P=|0\rangle\langle0|$ and $\dot P=0$. Pulsed measurements
give a survival probability $1-\Omega^2\delta t^2$ per step, hence a probability
decay rate $\Omega^2\delta t$. The absorber $H-i\kappa Q$ gives an amplitude
$\sim e^{-\Omega^2t/\kappa}$, hence a probability decay rate $2\Omega^2/\kappa$. 
Equating the two rates yields $\delta t=2/\kappa$, in agreement with
\eqref{eq:Schulman-SM}.

Since $\kappa$ is the decay rate of the amplitude in $\mathrm{Ran}\,Q$, it
is also the decay rate $\gamma_{\rm dec}$ of the coherence between the two
sectors, while probabilities decay at $2\kappa$. Equation
\eqref{eq:Schulman-SM} may therefore be recast in the convention-independent form
\begin{equation}
\delta t=\frac{2}{\gamma_{\rm dec}} ,
\label{eq:SchulmanUniversal-SM}
\end{equation}
which is Eq.~(14) of the main text. In this form the relation extends to
continuously monitored dynamics. As shown in Sec.~\ref{sec:Gcont}, homodyne
detection of $X=\sum_nx_nP_n(t)$ at rate $\kappa$ damps the coherence between
sectors $n$ and $m$ at $\gamma_{nm}=\tfrac{\kappa}{2}(x_n-x_m)^2$. Adiabatic
elimination of the off-diagonal blocks then yields the rate-weighted leakage operator
\eqref{eq:Gcont}. For a two-outcome PVM, where a single rate
$\gamma_{\rm dec}$ is available, this reduces to
$H_{\rm Z}-i\Gamma/\gamma_{\rm dec}$ with the very same $\Gamma$ as in
\eqref{eq:equality}, so that matching the three generators gives
$\delta t=2/\gamma_{\rm dec}=2/\kappa_{\rm CAP}$. For the choice $x_P=0$,
$x_Q=1$ one has $\gamma_{\rm dec}=\kappa/2$ and hence $\delta t=4/\kappa$,
illustrating that the naive identification $\delta t\leftrightarrow1/\kappa$ is
convention dependent whereas \eqref{eq:SchulmanUniversal-SM} is not.

We consider two limits as consistency checks. For a static projector ($\dot P=0$,
$A=0$), Eq.~\eqref{eq:equality} reduces to the standard static-Zeno leakage
$\Gamma=PHQHP$, establishing the model-independent generalization of the Schulman's original correspondence in the static case. For a
spectral projector of $H(t)$ (so that $QHP=0$), every term containing $QHP$ or
its adjoint $PHQ$ vanishes, leaving the purely geometric leakage
$\Gamma=P\dot P\dot P P$, controlled solely by the motion of the subspace. This
is the case used in the End Matter to derive the geometric bound on the
measurement rate.

Finally, we note that the opposite signs with which the geometric term enters the leading
generator and the leakage are consistent.  The
leading Kato--Avron term $+A$ originates in the frame derivative $iK$, whereas
the leakage coupling sees the co-moving image $W\tilde H W^\dagger=H-iK=H-A$ of
Eq.~\eqref{eq:WHW}. On the stroboscopic side the same combination $Q(H-A)P=
QHP-iQ\dot P P$ appears directly in Eq.~\eqref{eq:Gdt-square}, without reference
to any frame, which is why the two derivations agree.

\section{Continuous monitoring in the co-moving frame}\label{sec:SME}

We collect here the details of the co-moving treatment of the SME quoted in the
main text. Let $X(t)=\sum_nx_nP_n(t)$ be the monitored observable, with
nondegenerate eigenvalues $\{x_n\}$. In addition,  let $\rho_c$ obey the diffusive SME
[Eq.~(10) of the main text]
\begin{equation}
d\rho_c=-i[H(t),\rho_c]\,dt+\kappa\,\mathcal D[X(t)]\,\rho_c\,dt
+\sqrt{\kappa}\,\mathcal H[X(t)]\,\rho_c\,dW(t),
\label{eq:SME-SM}
\end{equation}
with $\mathcal D[X]\rho=X\rho X-\tfrac12\{X^2,\rho\}$,
$\mathcal H[X]\rho=\{X,\rho\}-2\langle X\rangle\rho$, and the associated
measurement record
\begin{equation}
dY(t)=2\sqrt{\kappa}\,\langle X(t)\rangle_c\,dt+dW(t),
\qquad
\langle\,\cdot\,\rangle_c=\mathrm{Tr}(\,\cdot\,\rho_c).
\end{equation}

\subsection{Inter-sector decoherence rates}

Averaging \eqref{eq:SME-SM} over the measurement outcomes, with $\mathbb E[dW]=0$, 
yields the unconditional master equation
$\dot\rho=-i[H,\rho]+\kappa\mathcal D[X]\rho$. Writing
$\rho_{nm}=P_n\rho P_m$ and using $XP_n=x_nP_n$, the dissipator acts on each
block as
\begin{equation}
\mathcal D[X]\rho_{nm}
=\Bigl(x_nx_m-\tfrac12x_n^2-\tfrac12x_m^2\Bigr)\rho_{nm}
=-\tfrac12(x_n-x_m)^2\rho_{nm},
\end{equation}
so that the coherence between sectors $n$ and $m$ is damped at the rate
\begin{equation}
\gamma_{nm}\;\equiv\;\gamma^{nm}_{\rm dec}=\frac{\kappa}{2}\,(x_n-x_m)^2.
\label{eq:gammadec}
\end{equation}
We use the shorthand $\gamma_{nm}$ below whenever the sector indices are
summed over.
The diagonal blocks ($n=m$) are unaffected, as they must be for a
quantum-nondemolition-type measurement of $X$. Equation \eqref{eq:gammadec} is
the quantity that enters the convention-independent Schulman relation, Eq. (14) in the main text.
%\eqref{eq:SchulmanUniversal-SM}.

\subsection{It\^o transformation to the co-moving frame}

Let $W(t)$ be the deterministic unitary intertwiner that transports the PVM,
\begin{equation}
\dot W(t)=K(t)\,W(t),
\qquad
K(t)\equiv-iA=\tfrac{1}{2}\sum_n[\dot P_n(t),P_n(t)],
\qquad W(0)=\mathbb I,
\label{eq:W-K-SM}
\end{equation}
for which $P_n(t)=W(t)P_n(0)W^\dagger(t)$ and hence $X(t)=W(t)X_0W^\dagger(t)$
with $X_0\equiv X(0)$. Define the co-moving conditioned state
$\tilde\rho_c(t)=W^\dagger(t)\rho_c(t)W(t)$. Because $W(t)$ is deterministic and
of bounded variation, the It\^o correction vanishes and
\begin{equation}
d\tilde\rho_c=W^\dagger(d\rho_c)W+(dW^\dagger)\rho_cW+W^\dagger\rho_c(dW)
=W^\dagger(d\rho_c)W-\bigl[K,\tilde\rho_c\bigr]dt .
\end{equation}
Substituting \eqref{eq:SME-SM} and using
$W^\dagger\mathcal D[X]\rho_c\,W=\mathcal D[X_0]\tilde\rho_c$ and
$W^\dagger\mathcal H[X]\rho_c\,W=\mathcal H[X_0]\tilde\rho_c$, one obtains an SME
of the same form but with time-independent measurement operators,
\begin{equation}
d\tilde\rho_c
=-i[\tilde H(t),\tilde\rho_c]\,dt
+\kappa\,\mathcal D[X_0]\,\tilde\rho_c\,dt
+\sqrt{\kappa}\,\mathcal H[X_0]\,\tilde\rho_c\,dW(t),
\label{eq:SME-tilde}
\end{equation}
with the transformed Hamiltonian
\begin{equation}
\tilde H(t)=W^\dagger(t)H(t)W(t)-i\,W^\dagger(t)\dot W(t),
\qquad
W^\dagger\dot W=W^\dagger\Bigl(\tfrac{1}{2}\sum_n[\dot P_n,P_n]\Bigr)W .
\label{eq:Htilde-SME}
\end{equation}
Thus, the same operator that generates the geometric term in the stroboscopic
derivation appears here as the inertial (connection) term $-iW^\dagger\dot W$ in
the co-moving frame; see also \cite{Lewalle2024}.

\subsection{Strong-measurement limit}

The Zeno regime corresponds to $\kappa$ being the largest rate in the problem.
In the co-moving frame \eqref{eq:SME-tilde}, the dissipator $\kappa\mathcal D[X_0]$
damps the off-diagonal blocks $P_n(0)\tilde\rho_cP_m(0)$, $n\neq m$, at the rates
\eqref{eq:gammadec}. As $\kappa\to\infty$, the dynamics becomes confined to the
block-diagonal manifold
$\tilde\rho_c(t)=\sum_nP_n(0)\tilde\rho_c(t)P_n(0)$,
and the remaining evolution within each block is generated, to leading order in
$1/\kappa$, by the projected Hamiltonian $P_n(0)\tilde H(t)P_n(0)$. Going
back to the laboratory frame with $H_{\rm eff}=iK+W(\sum_nP_n(0)\tilde HP_n(0))W^\dagger$
and using $\sum_nP_nAP_n=0$ reproduces Eq.~(5) of the main text,
\begin{equation}
H_{\rm Z}(t)=\sum_nP_n(t)H(t)P_n(t)+\tfrac{i}{2}\sum_n[\dot P_n(t),P_n(t)],
\end{equation}
in agreement with the stroboscopic Zeno result. If the monitored projectors are
the nondegenerate instantaneous spectral projectors of $H(t)$, the intertwiner
may be chosen as $W(t)=\sum_n|n(t)\rangle\langle n(0)|$ and the effective
Hamiltonian equals the CD Hamiltonian \eqref{eq:HZ-CD}. If instead one chooses
the two-outcome observable $X=P(t)$, the same analysis yields the Kato--Avron
Hamiltonian of Eq.~(4) of the main text. Carrying the elimination of the
off-diagonal blocks to next order yields, within each sector, a non-Hermitian
generator $H_{\rm Z}-i\Gamma_{\rm cont}$ built from the same combination $H-A$
as in the pulsed and absorptive routes, but with each leakage channel suppressed
by its own decoherence rate, as derived in Sec.~\ref{sec:Gcont}. For a 
two-outcome PVM it reduces to $H_{\rm Z}-i\Gamma/\gamma_{\rm dec}$, with $\Gamma$
the leakage operator \eqref{eq:equality}.

\subsection{Leakage operator under continuous monitoring}\label{sec:Gcont}

We now carry the elimination of the off-diagonal blocks to the next order and obtain
the leading non-Hermitian correction to the Zeno generator under continuous
monitoring, the analogue of Eq.~\eqref{eq:Gdt} for the stroboscopic protocol and
of Eq.~\eqref{eq:Gk} for the CAP protocol. Since the effective generator is a
property of the unconditional dynamics and independent of the unraveling, 
it suffices to work with the master equation obtained by averaging
\eqref{eq:SME-SM} over the measurement record,
$\dot\rho=-i[H,\rho]+\kappa\mathcal D[X]\rho$.

In the co-moving frame \eqref{eq:SME-tilde}, with $P_n^0\equiv P_n(0)$,
$\tilde\rho_{nm}\equiv P_n^0\tilde\rho\,P_m^0$ and
$\tilde H_{nm}\equiv P_n^0\tilde H P_m^0$, the dissipator is diagonal in the block
index and damps each coherence at its own rate \eqref{eq:gammadec},
\begin{equation}
\kappa\,\mathcal D[X_0]\,\tilde\rho_{nm}=-\gamma_{nm}\,\tilde\rho_{nm},
\qquad
\gamma_{nm}\equiv\frac{\kappa}{2}\,(x_n-x_m)^2 ,
\qquad \gamma_{nn}=0 .
\label{eq:gamma-nm}
\end{equation}
The equations of motion for the diagonal and off-diagonal blocks read
\begin{align}
\dot{\tilde\rho}_{nn}
&=-i\bigl[\tilde H_{nn},\tilde\rho_{nn}\bigr]
-i\sum_{l\neq n}\Bigl(\tilde H_{nl}\,\tilde\rho_{ln}-\tilde\rho_{nl}\,\tilde H_{ln}\Bigr),
\label{eq:diag-block}\\[2pt]
\dot{\tilde\rho}_{nm}
&=-\gamma_{nm}\,\tilde\rho_{nm}
-i\Bigl(\tilde H_{nm}\,\tilde\rho_{mm}-\tilde\rho_{nn}\,\tilde H_{nm}\Bigr)
+\mathcal O\bigl(\tilde H\,\tilde\rho_{\rm off}\bigr),
\qquad n\neq m .
\label{eq:offdiag-block}
\end{align}

In the Zeno regime the coherences are overdamped whenever
$\gamma_{nm}\gg\|\tilde H_{nm}\|$ for every pair $n\neq m$; they are then slaved
to the populations. Setting $\dot{\tilde\rho}_{nm}\simeq0$ in
\eqref{eq:offdiag-block},
\begin{equation}
\tilde\rho_{nm}\simeq-\frac{i}{\gamma_{nm}}
\Bigl(\tilde H_{nm}\,\tilde\rho_{mm}-\tilde\rho_{nn}\,\tilde H_{nm}\Bigr)
+\mathcal O(\gamma^{-2}) .
\label{eq:slaved}
\end{equation}
Substituting \eqref{eq:slaved} into \eqref{eq:diag-block}, the two terms of the
sum combine into a manifestly Lindbladian structure,
\begin{equation}
-i\Bigl(\tilde H_{nl}\tilde\rho_{ln}-\tilde\rho_{nl}\tilde H_{ln}\Bigr)
=\frac{1}{\gamma_{nl}}
\Bigl[\,2\,\tilde H_{nl}\,\tilde\rho_{ll}\,\tilde H_{ln}
-\bigl\{\tilde H_{nl}\tilde H_{ln},\,\tilde\rho_{nn}\bigr\}\Bigr],
\end{equation}
so that
\begin{equation}
\dot{\tilde\rho}_{nn}
=-i\bigl[\tilde H_{nn},\tilde\rho_{nn}\bigr]
+\sum_{l\neq n}\frac{1}{\gamma_{nl}}
\Bigl[\,2\,\tilde H_{nl}\,\tilde\rho_{ll}\,\tilde H_{ln}
-\bigl\{\tilde H_{nl}\tilde H_{ln},\,\tilde\rho_{nn}\bigr\}\Bigr]
+\mathcal O(\gamma^{-2}).
\label{eq:lindblad-blocks}
\end{equation}
Because $\gamma_{nl}$ is real and positive, the $\mathcal O(\gamma^{-1})$
correction is exclusively dissipative: no Hermitian (Lamb-type) shift is
generated at this order, in contrast with the case of a detuned, non-dissipative
elimination.

Going back to the laboratory frame with $H_{\rm eff}=iK+W(\cdot)W^\dagger$ and using
$W\tilde HW^\dagger=H-A$ from \eqref{eq:WHW} together with $\sum_nP_nAP_n=0$, the
diagonal generator becomes $H_{\rm Z}$ as in Sec.~\ref{sec:SME}. Further,  the
inter-sector terms of \eqref{eq:lindblad-blocks} map onto a Lindblad equation
with jump operators
\begin{equation}
L_{n\leftarrow l}=\sqrt{\frac{2}{\gamma_{nl}}}\;P_n(t)\bigl(H-A\bigr)P_l(t),
\qquad n\neq l ,
\label{eq:jump-ops}
\end{equation}
that transfer population from sector $l$ to sector $n$. The corresponding no-jump
(non-Hermitian) generator is $H_{\rm eff}=H_{\rm Z}-\tfrac{i}{2}\sum_{n\neq l}L^\dagger_{n\leftarrow l}L_{n\leftarrow l}$, that is,
\begin{equation}
H^{(\kappa)}_{\rm eff}(t)=H_{\rm Z}(t)-i\,\Gamma_{\rm cont}(t)+\mathcal O(\gamma^{-2}),
\qquad
\Gamma_{\rm cont}(t)=\sum_{n}\sum_{l\neq n}\frac{1}{\gamma_{nl}}\;
P_n\bigl(H-A\bigr)P_l\bigl(H-A\bigr)P_n 
\label{eq:Gcont}
\end{equation}
with $H_{\rm Z}=\sum_nP_nHP_n+A$ and $A=\tfrac{i}{2}\sum_n[\dot P_n,P_n]$. Since
$H$ and $A$ are Hermitian, each term in the sum is of the form $M^\dagger M$ with
$M=P_l(H-A)P_n$, so that
\begin{equation}
\Gamma_{\rm cont}
=\sum_{n\neq l}\frac{1}{\gamma_{nl}}\,
\bigl[P_l(H-A)P_n\bigr]^\dagger\bigl[P_l(H-A)P_n\bigr]\;\ge\;0.
\end{equation}
It follows that the continuous-monitoring leakage operator is positive, as in the
stroboscopic and CAP cases, and its anti-Hermitian contribution can only
deplete the norm within each sector.

Equation \eqref{eq:Gcont} shows that continuous monitoring is controlled by the
same combination $H-A$ of the physical Hamiltonian and the geometric
generator that governs the pulsed and CAP protocols, but that each leakage
channel $n\to l$ is suppressed by its own decoherence rate
$\gamma_{nl}\propto(x_n-x_l)^2$. The leakage is therefore weighted by the rate. For a
nondegenerate ladder of outcomes, $x_n=n$, one has
$\gamma_{nl}=\tfrac{\kappa}{2}(n-l)^2$ and transfer to distant sectors is
suppressed quartically in $|n-l|$. By contrast, in the stroboscopic and CAP protocols
 all channels are weighted equally. The three routes thus coincide, channel by channel, only
when all the relevant inter-sector rates are equal, in particular for the
two-outcome PVM used to fix the Schulman correspondence.

Let us next consider some limiting cases:

\emph{(i) Two-outcome PVM.---} For $\{P,Q\}$ with $x_P=0$, $x_Q=1$ there is a single
rate, $\gamma_{PQ}=\kappa/2\equiv\gamma_{\rm dec}$, and \eqref{eq:Gcont} collapses
onto the leakage operator \eqref{eq:equality} of the pulsed and CAP routes,
\begin{equation}
\Gamma_{\rm cont}=\frac{1}{\gamma_{\rm dec}}\,P\bigl(H-A\bigr)Q\bigl(H-A\bigr)P
=\frac{\Gamma}{\gamma_{\rm dec}},
\qquad
H^{(\kappa)}_{\rm eff}=H_{\rm Z}-\frac{i}{\gamma_{\rm dec}}\,\Gamma .
\label{eq:Gcont-two}
\end{equation}
Matching \eqref{eq:Gcont-two} with $H_{\rm Z}-\tfrac{i\delta t}{2}\Gamma$
returns $\delta t=2/\gamma_{\rm dec}$, Eq.~\eqref{eq:SchulmanUniversal-SM}, and
against the CAP generator $H_{\rm Z}-\tfrac{i}{\kappa_{\rm CAP}}\Gamma$ returns
$\kappa_{\rm CAP}=\gamma_{\rm dec}$. With $\gamma_{\rm dec}=\kappa/2$ this gives
$\delta t=4/\kappa$ for this choice of eigenvalues, confirming that the naive
identification $\delta t\leftrightarrow1/\kappa$ is convention dependent, whereas
the relation in terms of $\gamma_{\rm dec}$ is not.

\emph{(ii) Static PVM.---} For $\dot P_n=0$ one has $A=0$ and
$\Gamma_{\rm cont}=\sum_{n\neq l}\gamma_{nl}^{-1}P_nHP_lHP_n$, the familiar
Zeno-suppressed golden-rule rate, whose inverse dependence on the measurement
strength is the hallmark of the continuous quantum Zeno effect.

\emph{(iii) Spectral projectors.---} If $P_n(t)$ are the instantaneous
eigenprojectors of $H(t)$, then $P_nHP_l=0$ for $n\neq l$ and the leakage is
purely geometric,
\begin{equation}
\Gamma_{\rm cont}=\sum_{n\neq l}\frac{1}{\gamma_{nl}}\,
P_n\dot P_nP_l\dot P_nP_n ,
\end{equation}
so that the geometric bound of the End Matter applies channel by channel, with
$\gamma_{nl}$ playing the role of the inverse measurement interval.

As for the range of validity, 
the elimination requires $\gamma_{nl}\gg\|P_n(H-A)P_l\|$ for every pair
$n\neq l$, i.e.,\ $\kappa(x_n-x_l)^2\gg\|P_nHP_l\|,\,\|\dot P\|$. Outcomes that are
degenerate, $x_n=x_l$ with $n\neq l$, give $\gamma_{nl}=0$. That is,  they are not resolved
by the measurement and the corresponding sectors must be merged into a single
Zeno subspace before the elimination is performed, consistently with the
requirement of nondegenerate $\{x_n\}$ assumed in the main text.

\section{Multi-sector complex absorbing potentials}\label{sec:multiCAP}

The CAP construction of the main text, which distinguishes a single protected
subspace from its complement, generalizes to an arbitrary number of monitored
sectors. Consider the non-Hermitian Hamiltonian
\begin{equation}
H_{\rm nh}(t)=H(t)-i\kappa\,\Lambda(t),
\qquad
\Lambda(t)=\sum_n \lambda_n P_n(t),
\label{eq:multiCAP}
\end{equation}
with real coefficients $\lambda_n$ and $\kappa>0$ large. If all $\lambda_n$ are equal, then
$\Lambda(t)=\lambda\mathbb I$ and it induces only a global decay; therefore nontrivial sector separation requires
$\lambda_n\neq\lambda_m$ for some $n\neq m$.

In the co-moving frame $|\tilde\psi\rangle=W^\dagger|\psi\rangle$, with $W$ the
intertwiner \eqref{eq:W-K-SM},
the Schr\"odinger equation becomes
\begin{equation}
i\partial_t|\tilde\psi\rangle
=
\big(\tilde H(t)-i\kappa\Lambda_0\big)|\tilde\psi\rangle,
\qquad
\Lambda_0=\sum_n\lambda_n P_n(0),
\end{equation}
with $\tilde H(t)=W^\dagger H W-iW^\dagger\dot W$.
Decomposing $|\tilde\psi\rangle=\sum_n P_n(0)|\tilde\psi\rangle\equiv\sum_n|\tilde\psi_n\rangle$
yields
\begin{equation}
i\partial_t|\tilde\psi_n\rangle
=
\sum_m \tilde H_{nm}(t)|\tilde\psi_m\rangle
-i\kappa\lambda_n|\tilde\psi_n\rangle,
\end{equation}
where $\tilde H_{nm}=P_n(0)\tilde H P_m(0)$.
For $n\neq m$, transitions are suppressed when
$\kappa|\lambda_n-\lambda_m|
\gg\|\tilde H_{nm}(t)\|$.
Under this condition, adiabatic elimination of the off-diagonal sectors gives a leading-order effective generator
\begin{equation}
H_{\rm eff}(t)
=H_{\rm Z}(t)
-i\kappa\sum_n \lambda_n P_n(t)
+\mathcal O(\kappa^{-1}),
\end{equation}
which is block diagonal in the monitored decomposition. We note that this approach typically requires compensating gain in certain subspaces to preserve total probability.

A special case is the two-sector decomposition defined by a single projector
$P(t)$ and its complement $Q(t)=\mathbb I-P(t)$.
Choosing $\lambda_P=0$ and
$\lambda_Q=1$,
one recovers $H_{\rm nh}(t)=H(t)-i\kappa Q(t)$,
discussed in the main text.
In this case, the effective Zeno Hamiltonian becomes
$H_{\rm eff}(t)=P(t)H(t)P(t)+i[\dot P(t),P(t)]+\mathcal O(\kappa^{-1})$,
demonstrating that the multi-sector construction reduces to the
standard Kato--Avron Hamiltonian for a protected subspace, with the
$\mathcal O(\kappa^{-1})$ correction given by \eqref{eq:Gk}.

\end{document}